\documentclass[twocolumn]{aastex62}

\usepackage{graphicx}
\usepackage{times}
\usepackage{amssymb}
\usepackage{amsmath}
\usepackage{latexsym}
\usepackage[utf8]{inputenc}

\shorttitle{Chaos around black holes with discs or rings}
\shortauthors{L. Polcar, P. Sukov\'a, O. Semer\'ak}

\begin{document}

\title{Free motion around black holes with discs or rings:\\
       between integrability and chaos -- V}
\correspondingauthor{Oldrich Semer\'ak}
\email{oldrich.semerak@mff.cuni.cz}
\author{L. Polcar}
\affiliation{Institute of Theoretical Physics,
             Faculty of Mathematics and Physics,
             Charles University, Prague, Czechia}
\affiliation{Astronomical Institute,
             Czech Academy of Sciences,
             Ond\v{r}ejov, Czechia}
\author{P. Sukov\'a}
\affiliation{Astronomical Institute,
             Czech Academy of Sciences,
             Ond\v{r}ejov, Czechia}
\author{O. Semer\'ak}
\affiliation{Institute of Theoretical Physics,
             Faculty of Mathematics and Physics,
             Charles University, Prague, Czechia}

\begin{abstract}
Complete integrability of geodesic motion, the well known feature of the fields of isolated stationary black holes, can easily be ``spoilt'' by the presence of some additional source (even if highly symmetric). In previous papers, we used various methods to show how free time-like motion becomes chaotic if the gravitational field of the Schwarzschild black hole is perturbed by that of a circular disc or ring, considering specifically the inverted first disc of the Morgan-Morgan counter-rotating family and the Bach-Weyl ring as the additional sources. The present paper focuses on two new points. First, since the Bach-Weyl thin ring is physically quite unsatisfactory, we now repeat some of the analysis for a different, Majumdar-Papapetrou--type (extremally charged) ring around an extreme Reissner-Nordstr\"om black hole, and compare the results with those obtained before. We also argue that such a system is in fact more relevant astrophysically than it may seem. Second, we check numerically, for the latter system as well as for the Schwarzschild black hole encircled by the inverted Morgan-Morgan disc, how indicative is the geometric (curvature) criterion for chaos suggested by \cite{SotaSM-96}. We also add a review of the literature where the relevance of geometric criteria in general relativity (as well as elsewhere) has been discussed for decades.
\end{abstract}


\section{Introduction}

When studying a dynamical system, it goes without saying to ask how are the features of its evolution related to the properties of the governing interaction. Depending on the character of interaction, it is sometimes possible to interpret the dynamics as a geodesic flow in an effective manifold endowed with a metric obtained by a certain rescaling from the original, ``physical'' one \citep{Anosov-Sinai-67,Szydlowski-94}; the general-relativity theory itself stands for a celebrated example of such a ``geometrization'' of an interaction.\footnote
{Gravitation is a {\em universal} interaction, in particular, thanks to which it can be attributed to the underlying ``stage'' independently of which specific system is being studied, while still keeping the stage dynamical.}
If the latter has been achieved, a question naturally arises to which extent the features of the system's dynamics follow from quasi-local geometry of the obtained manifold. To say it straight away, a firm reply was given by \cite{VieiraL-96b}: while admitting that there surely exist links between global dynamics and local (curvature) properties of the corresponding configuration manifold of the system, they pointed out that ``any local analysis, in effective or even physical spaces, is far from being sufficient to predict a global phenomenon like chaotic motion'' (more recently, the same statement has e.g. been voiced by \cite{StranskyC-15}). One can easily imagine, for example, a system living in an Euclidean space and subject to an interaction only acting at discrete locations (and/or times), like e.g. collisions of a ball with an obstacle. In such a case, the (plane) geometry at generic location cannot alone tell anything about whether, when and with what result the ball is to hit a pole.

However, a general relativistic space-time is {\em not} discrete in the above sense, because any source (mass-energy) is being felt everywhere in it, not just at specific locations. More accurately, one should better restrict to stationary space-times, because, in a causal theory, changes are being ``announced'' with at most the speed of light, not immediately, so, for instance, a gravitational wave is in fact always ``surprising'' and cannot be anticipated from how the geometry at a given location looks ``now'' (the geometry can even be flat both before and after the wave). In a stationary case, the local geometry at any place and time is interconnected with how it looks elsewhere, so one may hope that when studying the evolution of some system in such a background, the local geometry could provide enough information for predicting the character of the system's global dynamics. This hope was strongly supported by observation \citep{Anosov-Sinai-67} that the geodesic flow is chaotic in manifolds whose all sectional curvatures are negative everywhere. Unfortunately, the result is only guaranteed for {\em compact manifolds without boundary}, while in real situations the accessible region often has some boundary which also affects the geodesic flow (there, the effective, Jacobi metric typically has a singularity).

Another reason to hope for the plausibility of geometric criteria in the study of a geodesic flow in space-time is that in this case one does not (or need not) refer to any effective manifold -- the geometry of the original, ``physical'' configuration space itself is relevant.
Within general relativity, the geometric criteria have notably been studied by Szyd{\l}owski and coworkers in the 1990s (see \citealt{Szydlowski-93,Biesiada-95,SzydlowskiK-96}, for example). \cite{Yurtsever-95} checked their reliability in a specific background of the Majumdar-Papapetrou--type binary of extremally charged static black holes (cf. also \citealt{Szydlowski-97}). He showed that null geodesics in this background can be treated as spatial geodesics on a certain two-dimensional Riemann surface and that the Gauss curvature of this surface is everywhere negative for any masses of the two black holes. The null geodesic flow can thus be expected to be chaotic, because negative curvature of a surface spanned by the flow is known to correspond to an exponential divergence of the geodesics (the flow has positive Lyapunov exponents). However, \cite{Yurtsever-95} asked then whether such properties (implying ``sensitive dependence on initial conditions'') are not only necessary, but also {\em sufficient} for the occurrence of chaos, and answered that {\em negatively}, pointing out that the flow must in addition be ``topologically mixing'' (mixing phase-space regions). He finished by demonstrating that the considered null geodesic flow does have the mixing property as well.

Soon after, \cite{SotaSM-96} studied the geodesic dynamics in several static axisymmetric space-times, namely those of the Zipoy-Vorhees class describing the fields of finite axial rods (and including e.g. the Schwarzschild and Curzon metrics as special cases), and also those generated by a system of Curzon-type point singularities distributed along the symmetry axis. The main focus was to compare standard methods like Poincar\'e maps, Lyapunov exponents or detection of a homoclinic tangle (which indicate chaos in the above situations) with criteria based on curvature properties of the space-time background. Specifically, they studied tidal-matrix eigen-values and curvature of the fictitious space obtained by energy-dependent conformal mapping within the Newtonian approach, while the eingenvalues of a matrix obtained from the Weyl tensor in the relativistic case.
\cite{VieiraL-96b} then demonstrated, however, that the curvature criterion suggested by \cite{SotaSM-96} is neither necessary nor sufficient for the occurrence of chaos, and showed that the disputed conclusion (about sufficiency of the proposed criterion) was probably made due to a wrong judgement about non-homoclinic character of a certain mode of chaotic behaviour that had been observed.

\cite{SzczesnyD-99} provided a useful summary of the reasoning based on the Maupertuis-Lagrange-Jacobi principle, connection between the interaction potential, the system's energy and the effective-metric curvature, the equation of geodesic deviation and the Lyapunov exponents of the flow. They demonstrated on examples how such a reasoning naturally provides a geometric criterion for local stability of the flow, but reminded that such a criterion could only be succesfull if the evolution were confined, within the effective Riemannian manifold, to a compact region {\em without} boundary.
Later \cite{Saa-04} illustrated the problem with the influence of the boundary on a system whose Jacobi-metric Gaussian curvature is everywhere positive in the accessible-region interior, yet which is still chaotic due to trajectories which ``bounce off'' the non-convex part of the boundary (like in famous billiard problems). \cite{SzydlowskiHS-96} suggested how to possibly circumvent this problem by releasing the smoothness of the manifold.

In the meantime, \cite{RamasubramanianS-01} (see also references therein) confirmed, on several Hamiltonian systems with 2 degrees of freedom, that the averaged sectional curvature defined by the effective-manifold geodesic congruence is closely related to the square of the averaged value of the maximal Lyapunov exponent, and conjectured a linear relation between these two quantities for Hamiltonian systems in general. Nevertheless, they pointed out that the above sectional curvature does {\em not} provide a sufficiently accurate indicator of the order-chaos transition.

For another direction of objections against generic reliability of ``geometric criteria for chaos'', see e.g. \cite{Wu-09} and references given therein: it has turned out that chaotic evolution may even be the case if the effective-manifold curvature is {\em positive}, but varying with position. The likely mechanism is the parametric instability due to some kind of resonance between the characteristics of the background curvature and those of the geodesic flow.

The purpose of the present paper is to further study the astrophysically motivated problem of motion of free test particles in static and axially symmetric fields generated by black holes surrounded by thin rings or discs. In previous papers, we used several different methods to reveal how the time-like geodesic flow becomes chaotic in dependence on parameters of the system (relative mass of the external source and particle energy, in particular): in \cite{SemerakS-10} we employed Poincar\'e sections, time series of position and velocity and their power spectra, and time evolution of the orbital ``latitudinal action''; in \cite{SemerakS-12} we classified the orbits according to the shape of the time-series power spectra, and also applying two recurrence methods, one based on tracing directions in which the trajectory recurrently passes through a pre-selected mesh of phase-space cells, and the other based on statistics over the recurrences themselves; in \cite{SukovaS-13} we computed several Lyapunov-type coefficients which quantify the rate of orbital divergence, namely the maximal Lyapunov characteristic exponent and the indicators called FLI and MEGNO (we specifically considered there a system involving a black hole surrounded by a small thin ``accretion'' disc or a large ring, having in mind the configuration which is observed in some galactic nuclei); finally, in \cite{WitzanySS-15} we compared the above exact, general relativistic treatment of the system with the Newtonian one, which mainly involved testing several ``pseudo-Newtonian'' potentials to mimic the central black hole.

In the present paper we check, on time-like geodesics in our black-hole--disc/ring field again, how indicative is the curvature criterion suggested by \cite{SotaSM-96}. Lead by a Newtonian case where local convergence/divergence of neighbouring trajectories is correlated with a sign of eigen-values of the tidal matrix (given by second derivatives of the gravitational potential), \cite{SotaSM-96} followed the above quoted results by Szyd{\l}owski et al. and considered a local criterion based on certain eigen-values of the Riemann tensor. After a short summary of the criterion in Sect. \ref{Sota-criterion}, we compare its guess with numerical study of the actual geodesic flow in Sect. \ref{Sota-criterion,numerics}.

Besides the above aim, we also include one novelty, namely, when considering the thin ring placed around a black hole in a concentric way, we choose a different solution than in previous papers of this series. There, we took the aged Bach-Weyl solution \cite{Bach-Weyl-22} which is being considered the default solution of this type since it corresponds to a Newtonian homogeneous-ring solution for the potential ($\nu$), supplemented by an appropriate second metric function $\lambda$ according to the field equations. However, we showed in \cite{Semerak-16} that the Bach-Weyl (BW) ring is a rather strange source, mainly in that its field is not locally cylindrical (the ring is infinitely remote ``from within'', whereas at finite distance ``from outside''). We compared it with several other ring-singularity solutions there and found, in particular, that physically much more reasonable (not directional, in particular) is the field generated by what we called there the Majumdar-Papapetrou (MP) ring, namely a homogeneous circular thin ring bearing an extremal value of electric charge (its density equals the mass density). Thanks to an exact balance between gravitational attraction and electrostatic repulsion, solutions of this kind are known to provide one of rare options for multi-body equilibria (and even {\em the only} option for equilibria of multiple black holes).
Within the class of the Majumdar-Papapetrou solutions the superposition is very simple, in particular, the function $\lambda$ remains zero, so we also choose the {\em extreme Reissner-Nordstr\"om} black hole (which is the MP-type source as well) instead of the Schwarzschild one as the central body.

Note that we use geometrized units in which $c\!=\!1$ and $G\!=\!1$ and metric signature ($-$$+$$+$$+$). The convention for the Riemann curvature tensor ${R^\mu}_{\nu\kappa\lambda}$ is given by Ricci identity for commutator of covariant derivatives, $V_{\nu;\kappa\lambda}\!-\!V_{\nu;\lambda\kappa}={R^\mu}_{\nu\kappa\lambda}V_\mu$. Greek indices represent space-time components (0, 1, 2, 3). We set the cosmological constant to zero. Let us also stress here that we study just geodesic motion in the above space-times, that is, the test particles only interact with the disc gravitationally (if crossing the disc, they do not interact mechanically, in particular).

\subsection{Note on extremally charged sources}

Since one always stresses that extremally charged sources have zero astrophysical relevance, we should more explain why we have included them in this study. As already mentioned above, the main reason is that the exact general relativistic gravitational field (and mainly curvature) behaves much more reasonably in the vicinity of the Majumdar-Papapetrou ring than around the Bach-Weyl ring \citep{Semerak-16}. Actually, though the Bach-Weyl ring is a straightforward counter-part of the Newtonian circular ring, it turns out to be a directional singularity -- it behaves differently when approached from different local meridional directions, even lying at infinite proper distance when approached from within. Besides being thus rather unacceptable as a model of an astrophysical ring (thin toroid), we also suspected that it is just this weird property what might induce geodesic chaos in our system, so we needed to employ some more reasonable ring source to verify this -- and the Majumdar-Papapetrou--type ring is the only other static circular thin ring which is available. (We focus on comparison of the two configurations in Sect. \ref{Schw+BW,RN+MP}.)

Needless to say, we are not saying that the astrophysical sources are significantly charged. However, being charged is in fact a good rather than bad feature of the configuration we newly included, because the electromagnetic field which it brings along may {\em mimic} the gravitational effect of diluted matter in the system. Namely, real accreting black-hole systems are not vacuum, but rather immersed in gas. Although the electromagnetic field is of course a different source than gas, it has the same main property -- a positive mass-energy density. And this property is in fact the only important one in our case, because we are interested in the motion of {\em uncharged} particles which only feel the gravitational effect of the sources (including that of their electromagnetic field); they do not feel the Lorentz force. One can add that the electromagnetic-field energy density even behaves in a desirable way in that it is maximal at the sources (black hole and the ring) and falls off when receding from them, like what one could expect for a gaseous environment. More specifically, for the Majumdar-Papapetrou family of solutions, the electrostatic potential $\Phi$ is related to the gravitational potential $\nu$ in a very simple way, $\Phi=e^\nu+{\rm const}$.

To summarize this point, it is because of physical reasons, not just to study ``yet another source'', that we consider charged sources here. We are far from claiming that accreting black holes are properly described by extremally charged configurations, we only claim that the configuration we have newly included here can describe the {\em gravitational} field of accreting black holes more adequately than the superposition of Schwarzschild with the Bach-Weyl ring which we used in previous papers of this series.

\section{Geometric criterion based on eigen-values of the Riemann tensor}
\label{Sota-criterion}

Following \cite{SotaSM-96}, let us rewrite the geodesic deviation equation for a tangent field $u^\mu$ and a transversal relative position $n^\mu$ (supposed to be orthogonal, $u_\mu n^\nu\!=\!0$) as
\begin{align}
  \frac{{\rm D}^2 n^\mu}{{\rm d}\tau^2}
  &= -{R^\mu}_{\nu\kappa\lambda}u^\nu n^\kappa u^\lambda =  \nonumber \\
  &= -\frac{1}{2}\,g^{\mu\nu}\,\frac{\partial}{\partial n^\nu}
      (R_{\alpha\beta\gamma\delta}n^\alpha u^\beta n^\gamma u^\delta) =  \nonumber \\
  &= -\frac{1}{2}\,g^{\mu\nu}\,\frac{\partial}{\partial n^\nu}
      (E^\alpha_\gamma n_\alpha n^\gamma) \,,
\end{align}
where $E^\alpha_\gamma\!:=\!{R^{\alpha\beta}}_{\gamma\delta}u_\beta u^\delta$ is the ``electric'' part of the Riemann/Weyl tensor (in the frame tied to $u^\mu$) and ${\rm D}/{\rm d}\tau$ is the absolute derivative along $u^\mu$, with $\tau$ denoting proper time. The expression $\frac{1}{2}E^\alpha_\gamma n_\alpha n^\gamma$ clearly represents a kind of tidal-force potential. This potential has a saddle point (with zero value) at $n^\mu\!=\!0$ and its behaviour in the vicinity of this point tells in which $n^\mu$ directions the geodesics converge/diverge. Which of the two tendencies prevails can be learned from eigen-values of $E^\alpha_\gamma$.

Because the energy-momentum tensor of an electromagnetic field is traceless and we do not consider any other source (and assume zero cosmological constant), the Einstein equations imply zero Ricci scalar and this in turn means that the tensor $E^\alpha_\gamma$ is traceless in a vacuum.
As a 4$\times$4 matrix, $E^\alpha_\gamma$ has four eigen-values; the sum of these is zero due to the zero trace (in the vacuum). Also zero is one of the eigen-values, namely that corresponding to the eigen-direction given by $u^\gamma$,
\[E^\alpha_\gamma u^\gamma = {R^{\alpha\beta}}_{\gamma\delta}u_\beta u^\gamma u^\delta = 0 \,.\]
However, $n^\gamma$ never belongs to this eigen-subspace, because it is orthogonal to $u^\gamma$. Hence, relevant are the remaining three eigen-values.

The electric part of the Riemann tensor is dependent on the four-velocity $u^\mu$ which however is not a priori known (there is no analytical formula for it). Nevertheless, it has been shown (see \citealt{SotaSM-96}) that the non-zero eigen-values of $E^\alpha_\gamma$ as well solve the related eigen-problem
\begin{equation}
  R^A_B S^B=\kappa S^A \,,
\end{equation}
where $S^A$ is a column vector associated with the bivector $S^{\mu\nu}=n^\mu u^\nu-u^\mu n^\nu$ and $R^A_B$ is a 6x6 matrix which represents the action of the Riemann tensor on the space of bivectors. The matrix $R^A_B$ has in total six eigen-values, and fixing the $u^\mu$ in $S^{\mu\nu}$ exactly selects those three of them which are also the eigen-values of $E^\alpha_\gamma$.

Consider now the Weyl metric
\begin{equation}  \label{Weyl-metric}
  {\rm d}s^2=-N^2{\rm d}t^2+N^{-2}\left[\rho^2{\rm d}\phi^2+e^{2\lambda}({\rm d}\rho^2+{\rm d}z^2)\right],
\end{equation}
where $t$ and $\phi$ are Killing coordinates (on which the metric does not depend), so the two metric functions $N\!\equiv\!e^\nu$ (lapse) and $\lambda$ only depend on $\rho$ and $z$ which cover (isotropically) the meridional plane. Note that the static and axially symmetric electro-vacuum region can always be described in this form.

For the Weyl metric (\ref{Weyl-metric}), the curvature matrix $R^A_B$ assumes a simple block-diagonal form ${\rm diag}(R_1,R_2)$, where $R_1$ and $R_2$ are 3x3 matrices. For electro-vacuum space-times, $R^A_B$ is traceless and thus its six eigen-values sum to zero. For {\em vacuum} space-times, $R_1\!=\!R_2$ in addition, so we are only left with three eigen-values which can be expressed as
\begin{align}
  \kappa_1 &=
  \frac{1}{2}\left[{R^{t\rho}}_{t\rho}+{R^{tz}}_{tz}+
                   \sqrt{({R^{t\rho}}_{t\rho}\!-\!{R^{tz}}_{tz})^2+4({R^{t\rho}}_{tz})^2}\right], \\
  \kappa_2 &=
  \frac{1}{2}\left[{R^{t\rho}}_{t\rho}+{R^{tz}}_{tz}-
                   \sqrt{({R^{t\rho}}_{t\rho}\!-\!{R^{tz}}_{tz})^2+4({R^{t\rho}}_{tz})^2}\right], \\
  \kappa_3 &= {R^{t\phi}}_{t\phi}={R^{\rho z}}_{\rho z}
\end{align}
(so they are independent of $u^\mu$).
Specifically, they are, together with the respective eigen-vectors $n^\gamma$, solutions of the equation
\[E^\alpha_\gamma n^\gamma=-\kappa\,n^\alpha.\]

For the vacuum case, the range of possibilities is quite restricted. Since the eigen-values add up to zero, they cannot all be of the same sign. Recalling the original equation of geodesic deviation, it can be expected that the geodesics should diverge in regions where two of the eigen-values are positive, whereas convergence should prevail where two of them are negative. As clearly $\kappa_1\!>\!\kappa_2$, it is not possible that $\kappa_1\!<\!0$ and $\kappa_2\!>\!0$, so one can specify that there are just two ``diverging'' cases of $(\kappa_1,\kappa_2,\kappa_3)$: ($+$$+$$-$) and ($+$$-$$+$). \cite{SotaSM-96} conjectured -- and also illustrated on examples -- that the ($+$$+$$-$) region can be expected to induce more instability, because there the geodesics diverge within the meridional, $(\rho,z)$ plane, whereas in the ($+$$-$$+$) region the divergence is restricted solely to Killing directions, which should not be much important for the tendency to chaos.

For the case with non-zero electromagnetic field, like for the Majumdar-Papapetrou space-times, the submatrices $R_1$ and $R_2$ differ, but we have checked that in our case the eigen-values of {\em each} of these submatrices almost exactly sum to zero, and that the unstable regions determined by their respective sets of eigen-values almost exactly coincide (in pairs). Therefore, it is in fact sufficient to analyse just the three eigen-values of $R_1$ (for example) and proceed similarly as in the vacuum (there again exist just two types of unstable regions).

In order to test the criterion, we thus localize, for a given metric and geodesics with given constants of motion, the potentially unstable regions (according to the above eigen-values), check whether and ``how much'' of the phase-space volume accessible to the geodesics they occupy (and so can affect the latter), and finally compare the resulting prediction with numerical integration of geodesics (visualized, e.g., on Poincar\'e diagrams). Let us recall here that the accessible region is given by normalization of four-velocity $g_{\mu\nu}u^\mu u^\nu=-1$ which for our metric (\ref{Weyl-metric}) can be rewritten as
\begin{equation}
  e^{2\lambda}\left[(u^\rho)^2+(u^z)^2\right]={\cal E}^2-N^2(1+N^2\ell^2/\rho^2) \,,
\end{equation}
where ${\cal E}\!:=\!u_t$ and $\ell\!:=\!u_\phi$ are the specific energy and azimuthal angular momentum with respect to infinity, conserved along any geodesic. Clearly the right-hand side has to be non-negative and this fixes the region where the geodesics with given ${\cal E}$ and $\ell$ are bound to.

\section{Extreme Reissner-Nordstr\"om (RN) black hole and the Majumdar-Papapetrou (MP) ring}

We will check how the geometric criterion works for a time-like geodesic flow in the static and axisymmetric background of an extreme RN black hole surrounded, in a concentric manner, by a homogeneous circular thin ring also bearing an extremal charge (its charge density is the same as the mass density). Both sources belong to a wider class of MP solutions, a subclass of Weyl space-times\footnote
{The basic properties of the whole Weyl class of space-times were given in previous papers of this series, mainly in the first one, so we do not repeat them here.}

with properties fully represented by just one function, the lapse $N$, the second Weyl-metric function $\lambda$ being zero. Hence the Weyl form of the MP metric
\begin{equation}
  {\rm d}s^2 = -N^2{\rm d}t^2+N^{-2}\left(\rho^2{\rm d}\phi^2+{\rm d}\rho^2+{\rm d}z^2\right).
\end{equation}

For the extreme RN black hole, the lapse function is given by
\begin{equation}
  \frac{1}{N} = 1+\frac{M}{\sqrt{\rho^2+z^2}} \;,
\end{equation}
while for the MP ring it is given by
\begin{align}
  \frac{1}{N}
  & =1+\frac{{\cal M}}{2\pi}\int_0^{2\pi}\frac{{\rm d}\phi'}{\sqrt{\rho^2+b^2-2b\rho\cos(\phi\!-\!\phi')+z^2}}=
  \nonumber \\
  & =1+\frac{2{\cal M}K(k)}{\pi l_2} \;,  \label{1/N}
\end{align}
where $b$ stands for the ring's Weyl radius and ${\cal M}$ for its total mass (we write it in calligraphic to distinguish it from the black-hole mass $M$),
\[l_{1,2}:=\sqrt{(\rho\mp b)^2+z^2}\]
and
\[K(k) := \int_0^{\pi/2}\frac{{\rm d}\alpha}{\sqrt{1-k^2\sin^2\alpha}}\]
is the complete elliptic integral of the first kind, with modulus and complementary modulus
\[k^2:=1-\frac{(l_1)^2}{(l_2)^2}=\frac{4b\rho}{(l_2)^2}\;, \qquad
  k'^2:=1-k^2=\frac{(l_1)^2}{(l_2)^2} \;.\]
Note that the second term of (\ref{1/N}) represents the (minus) Newtonian potential of a homogeneous infinitesimally thin ring.

The basic properties of the MP-ring space-time were studied in \cite{Semerak-16}. We showed there that it has quite reasonable properties in comparison with other static axisymmetric ring solutions, mainly in comparison with the BW ring. Most notably and in contrast to the BW ring, the MP ring is locally cylindrical, so it is same from all ``local latitudinal'' directions. Actually, like for every Weyl-type solution, the gravitational potential $\nu$ (thus lapse) can be taken over from Newtonian treatment (because it is given by the same Laplace equation in both cases) and no surprise occurs at this level, but the second metric function $\lambda$ (which has no Newtonian counter-part) can bring serious deformation to the space-time; the BW ring is an example. For the MP ring, on the contrary, $\lambda\!=\!0$, so the field is completely represented just by the potential $\nu$ itself and there is no additional deformation.

\subsection{Superposition of the sources}

For a general Weyl-type solution, the Einstein equations imply that the potential $\nu$ has to satisfy
\begin{equation}
  \nu_{,\rho\rho}+\frac{\nu_{,\rho}}{\rho}+\nu_{,zz}
    = e^{-2\nu}\left[(\Phi_{,\rho})^2+(\Phi_{,z})^2\right],
\end{equation}
where $\Phi$ is the electrostatic potential.
For the MP subclass of solutions, this potential is related to the gravitational potential $\nu$ by
\[\Phi=e^\nu+{\rm const},\]
so the above equation becomes
\begin{equation}
  \nu_{,\rho\rho}+\frac{\nu_{,\rho}}{\rho}+\nu_{,zz} = (\nu_{,\rho})^2+(\nu_{,z})^2.
\end{equation}
This is equivalent to the equation
\begin{equation}
  N_{,\rho\rho}+\frac{N_{,\rho}}{\rho}+N_{,zz}
   = \frac{2}{N}\left[(N_{,\rho})^2+(N_{,z})^2\right]
\end{equation}
for $N\!\equiv\!e^\nu$ and can be rewritten as a Laplace equation for $1/N$,
\begin{equation}
  (1/N)_{,\rho\rho}+\frac{1}{\rho}(1/N)_{,\rho}+(1/N)_{,zz} = 0 \,.
\end{equation}
Hence, adding two solutions of the MP type means to superpose linearly their $1/N$. Together with the requirement that $N\!\rightarrow\!1$ at spatial infinity, the superposition of the extreme RN black hole with the MP ring is thus described by the total value
\begin{equation}
  \frac{1}{N} = 1+\frac{M}{\sqrt{\rho^2+z^2}}+\frac{2{\cal M}K(k)}{\pi l_2} \;.
\end{equation}

\section{Geodesic dynamics in the field of the RN black hole encircled by the MP ring}
\label{Schw+BW,RN+MP}

\begin{figure*}
\includegraphics[width=\textwidth]{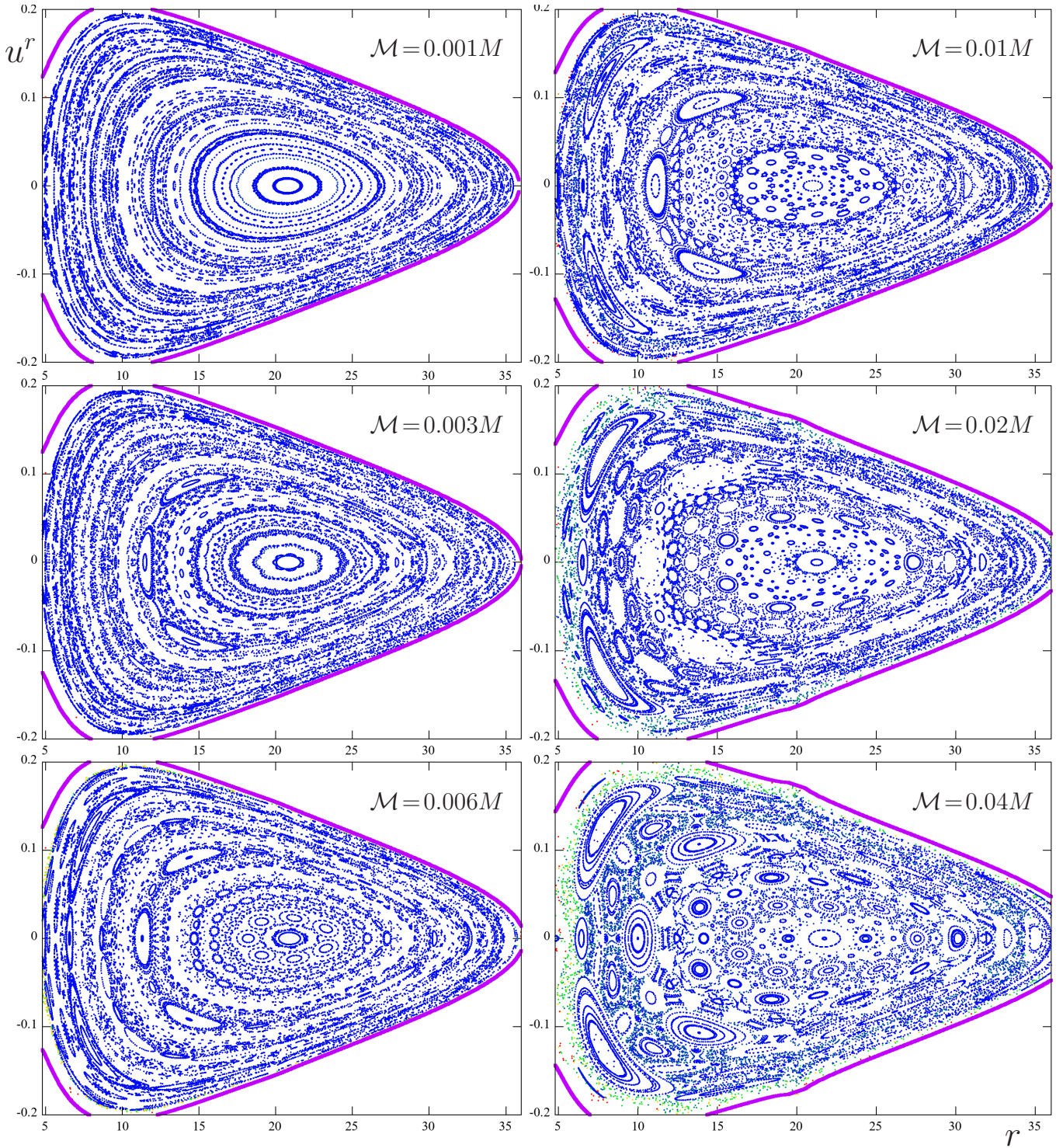}
\caption
{Equatorial Poincar\'e sections for geodesics in the field of a Schwarzschild black hole encircled by a Bach-Weyl ring with radius $r_{\rm ring}\!=\!20M$: dependence on relative ring mass ${\cal M}/M$ (its value is indicated in the plots). Passages of orbits having $\ell\!=\!3.75M$, ${\cal E}\!=\!0.977$ through the ring plane are drawn. See the main text for description of the colouring. Figure continues on the next page.}
\label{Schw-BW-mass}
\end{figure*}

\begin{figure*}
\includegraphics[width=\textwidth]{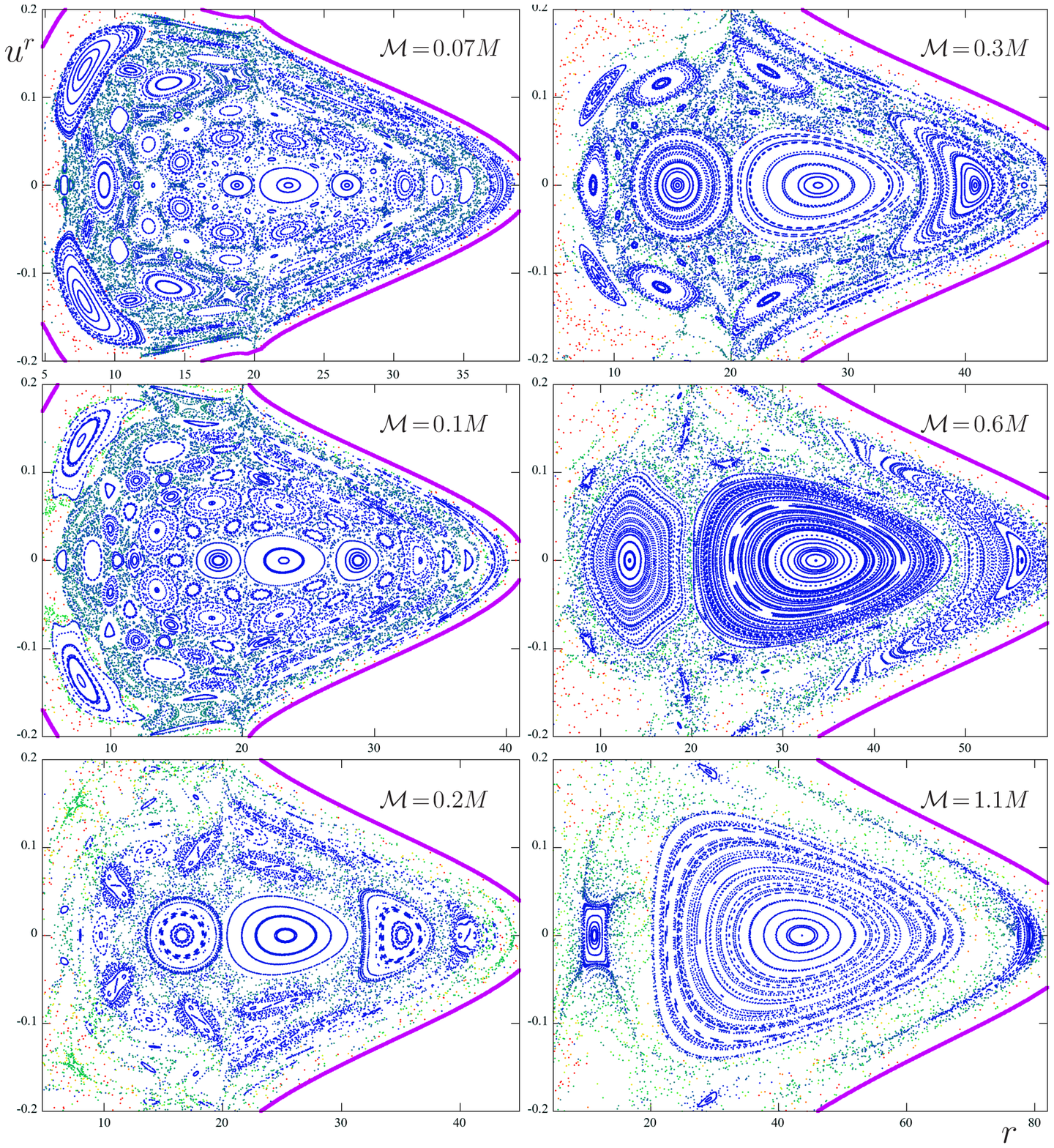}
\end{figure*}

\begin{figure*}
\includegraphics[width=\textwidth]{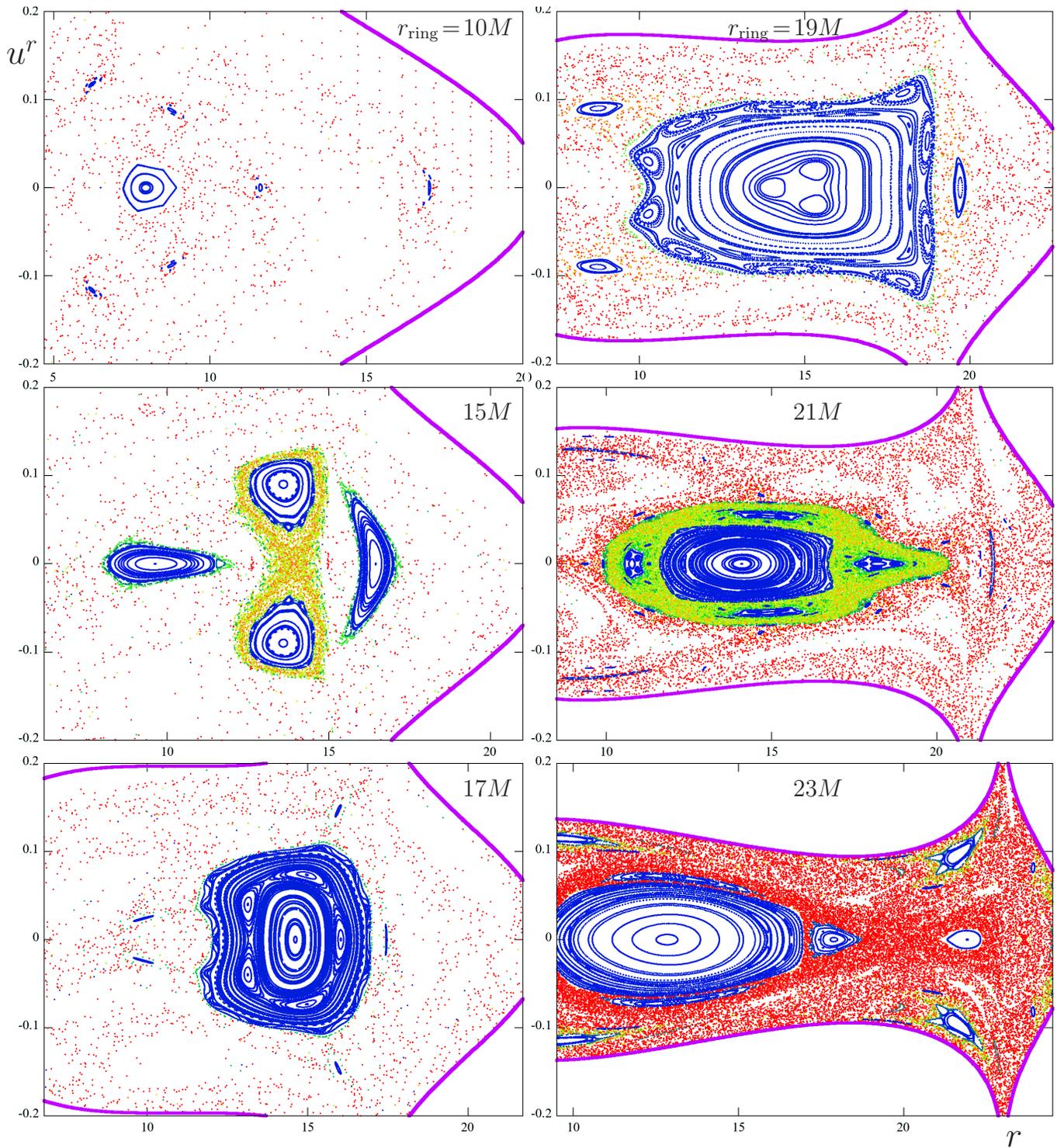}
\caption
{Similar type of Poincar\'e diagrams as in Fig. \ref{Schw-BW-mass}, now showing dependence of geodesic dynamics on radius of the BW ring $r_{\rm ring}$ (its values are again given in the plots). The ring mass is set at ${\cal M}\!=\!0.5M$ and the geodesics have constants of motion $\ell\!=\!3.75M$, ${\cal E}\!=\!0.94$.}
\label{Schw-BW-radius}
\end{figure*}

\begin{figure*}
\includegraphics[width=\textwidth]{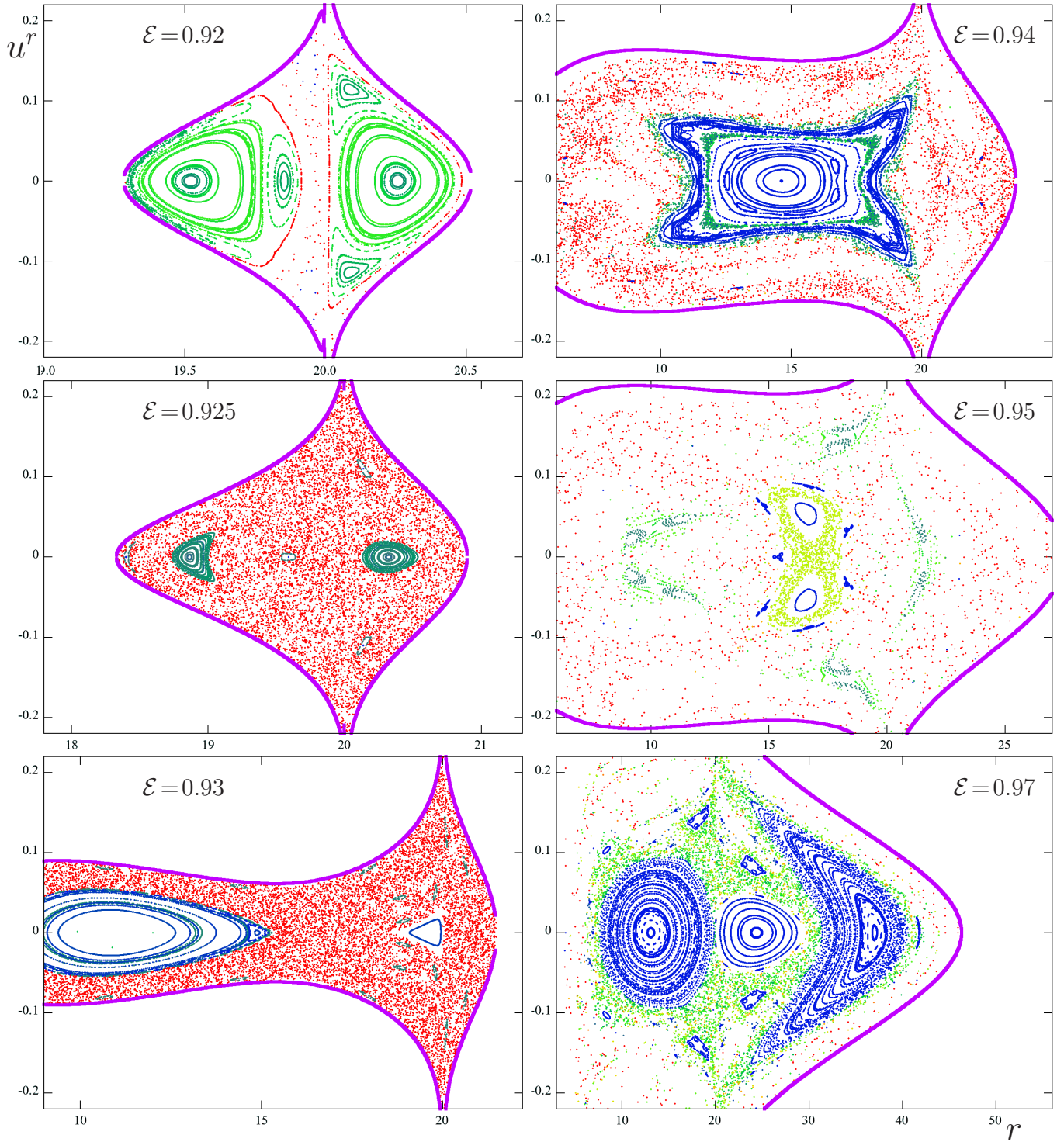}
\caption
{Similar type of Poincar\'e diagrams as in previous two figures, now showing dependence on the specific energy of geodesics ${\cal E}$ (its values are again given in the plots). The ring mass and radius are ${\cal M}\!=\!0.5M$, $r_{\rm ring}\!=\!20M$, and geodesics have angular momentum $\ell\!=\!3.75M$.}
\label{Schw-BW-energy}
\end{figure*}

\begin{figure*}
\includegraphics[width=\textwidth]{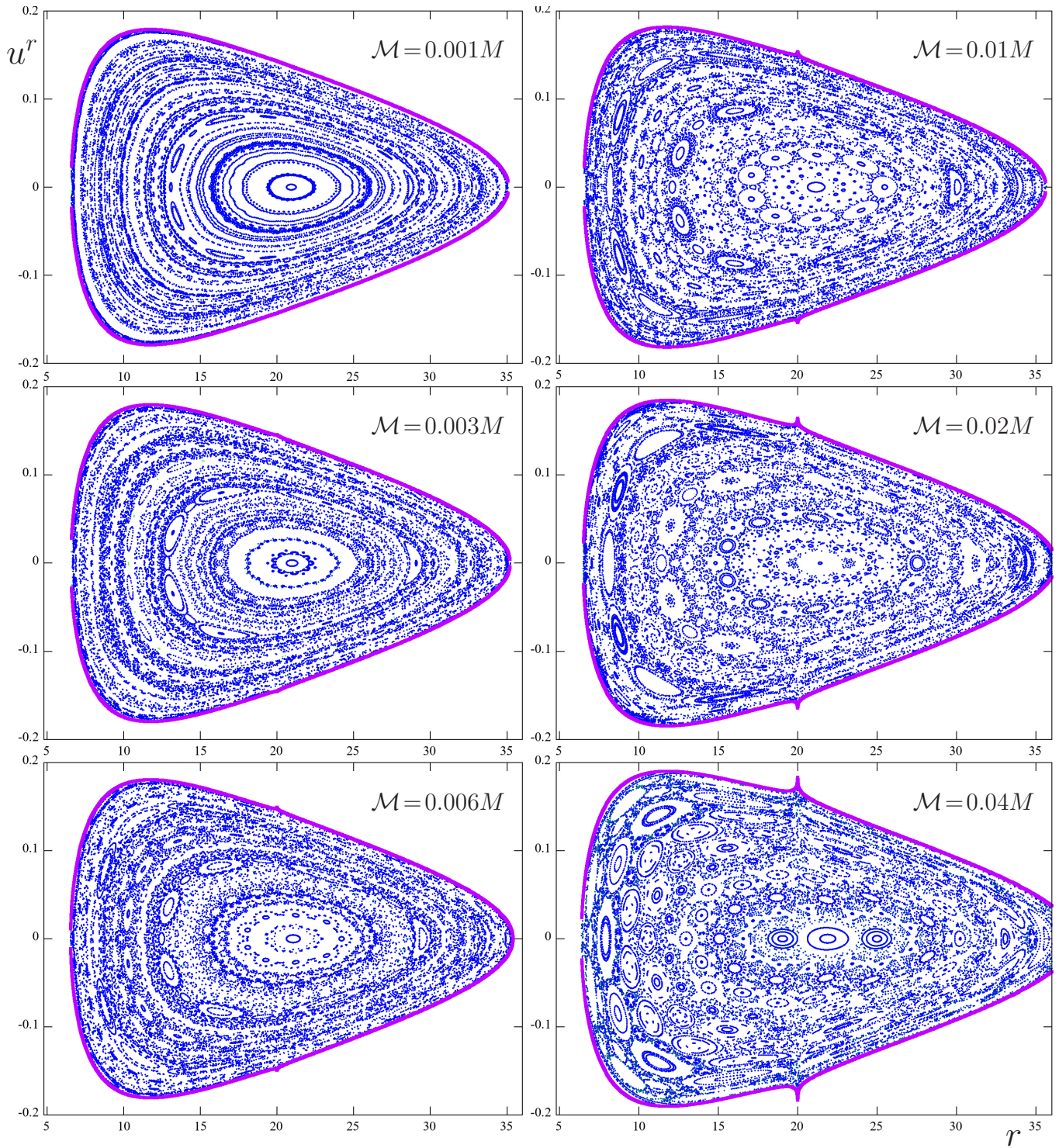}
\caption
{Counterpart of Fig. \ref{Schw-BW-mass} obtained for geodesics in the field of an extreme Reissner-Nordstr\"om black hole encircled by a Majumdar-Papapetrou ring. Parameters $r_{\rm ring}$, $\ell$ and ${\cal E}$ (and also ${\cal M}$, as indicated in the plots) are chosen the same as in the Schwarzschild + Bach-Weyl ring case. Figure continues on the next page.}
\label{RN-MP-mass}
\end{figure*}

\begin{figure*}
\includegraphics[width=\textwidth]{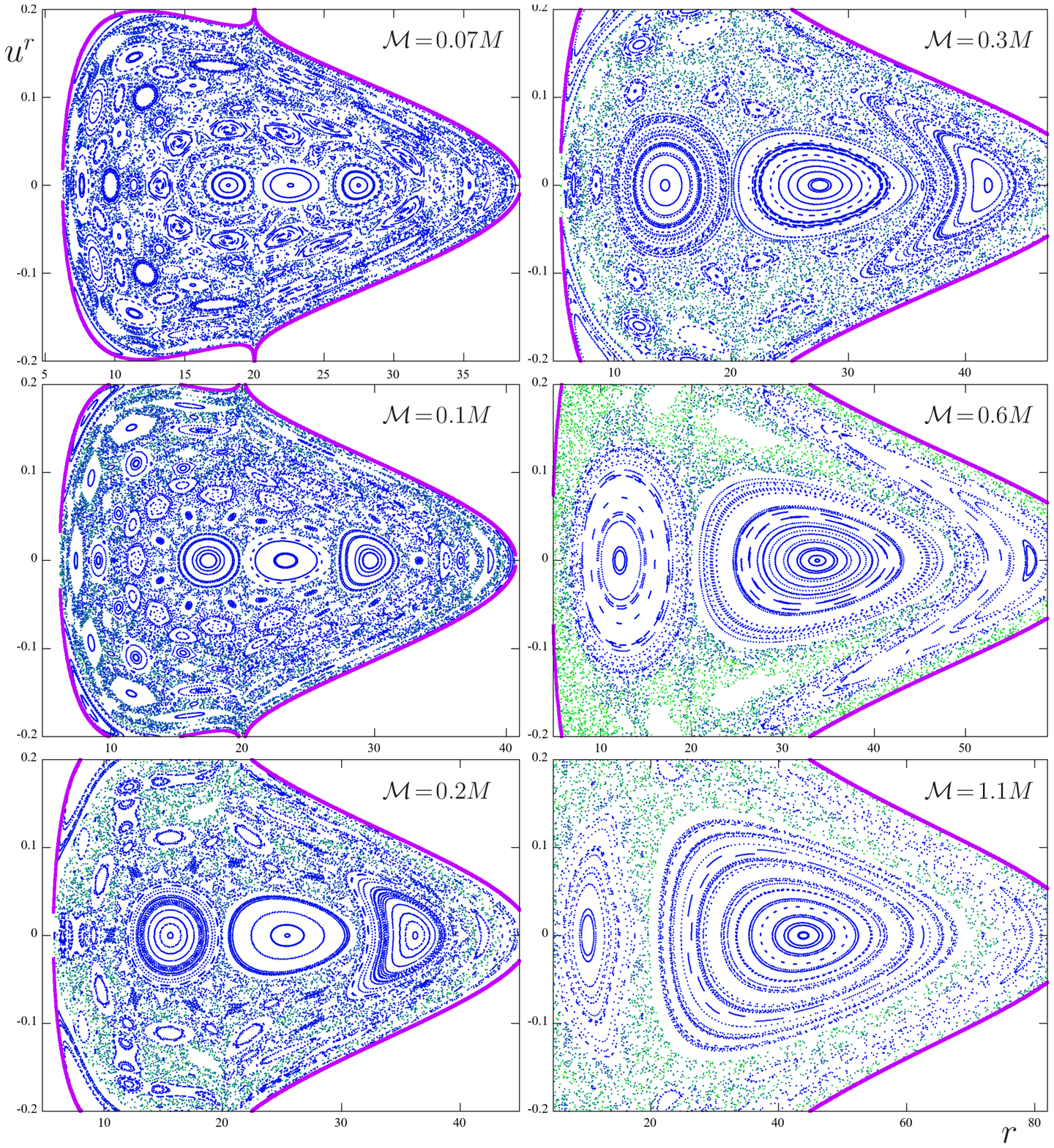}
\end{figure*}

\begin{figure*}
\includegraphics[width=\textwidth]{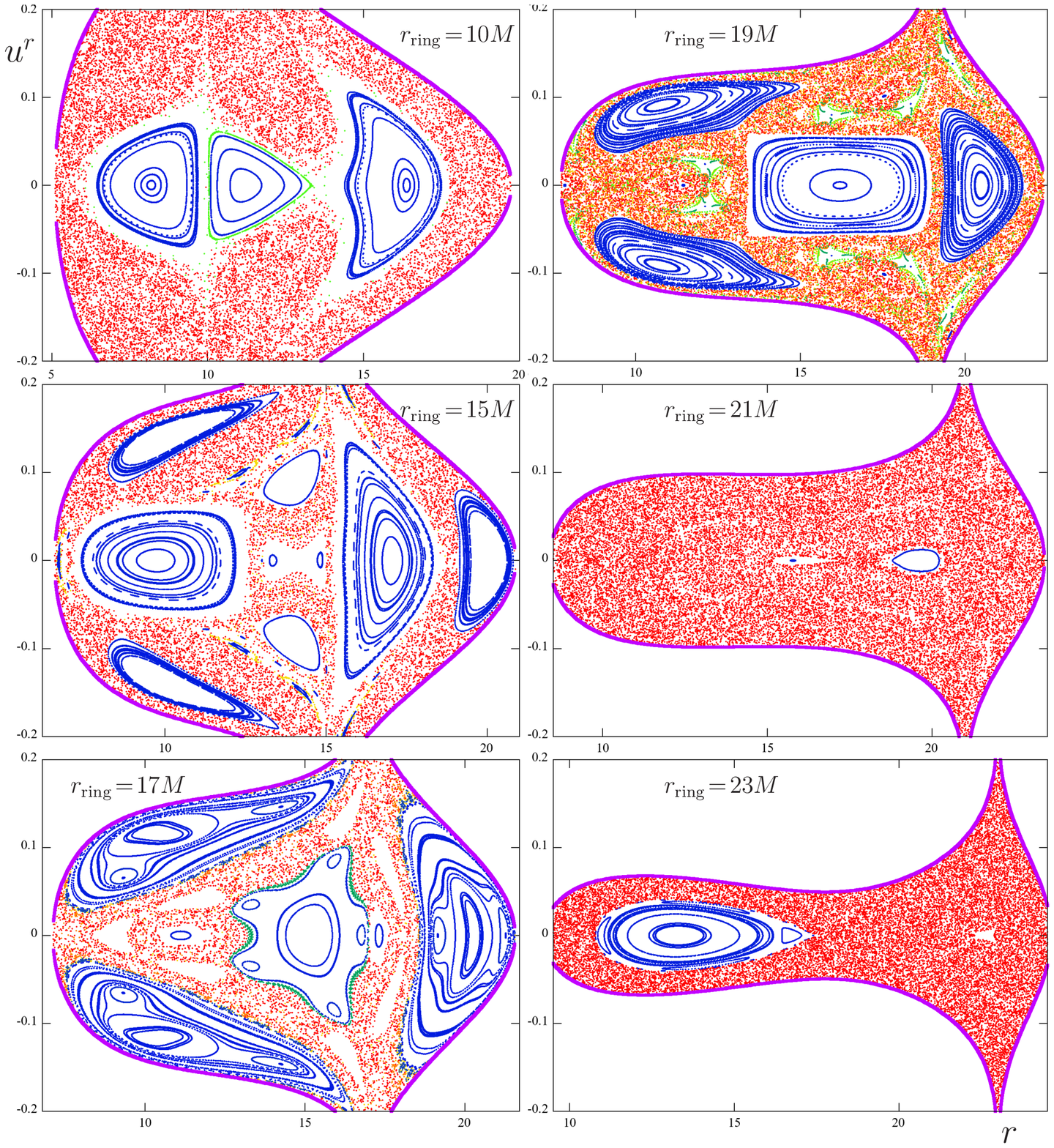}
\caption
{Counterpart of Fig. \ref{Schw-BW-radius} obtained for geodesics in the field of an extreme Reissner-Nordstr\"om black hole encircled by a Majumdar-Papapetrou ring. Parameters ${\cal M}$, $\ell$ and ${\cal E}$ (and also $r_{\rm ring}$, as indicated in the plots) are chosen the same as in the Schwarzschild + Bach-Weyl ring case.}
\label{RN-MP-radius}
\end{figure*}

\begin{figure*}
\includegraphics[width=\textwidth]{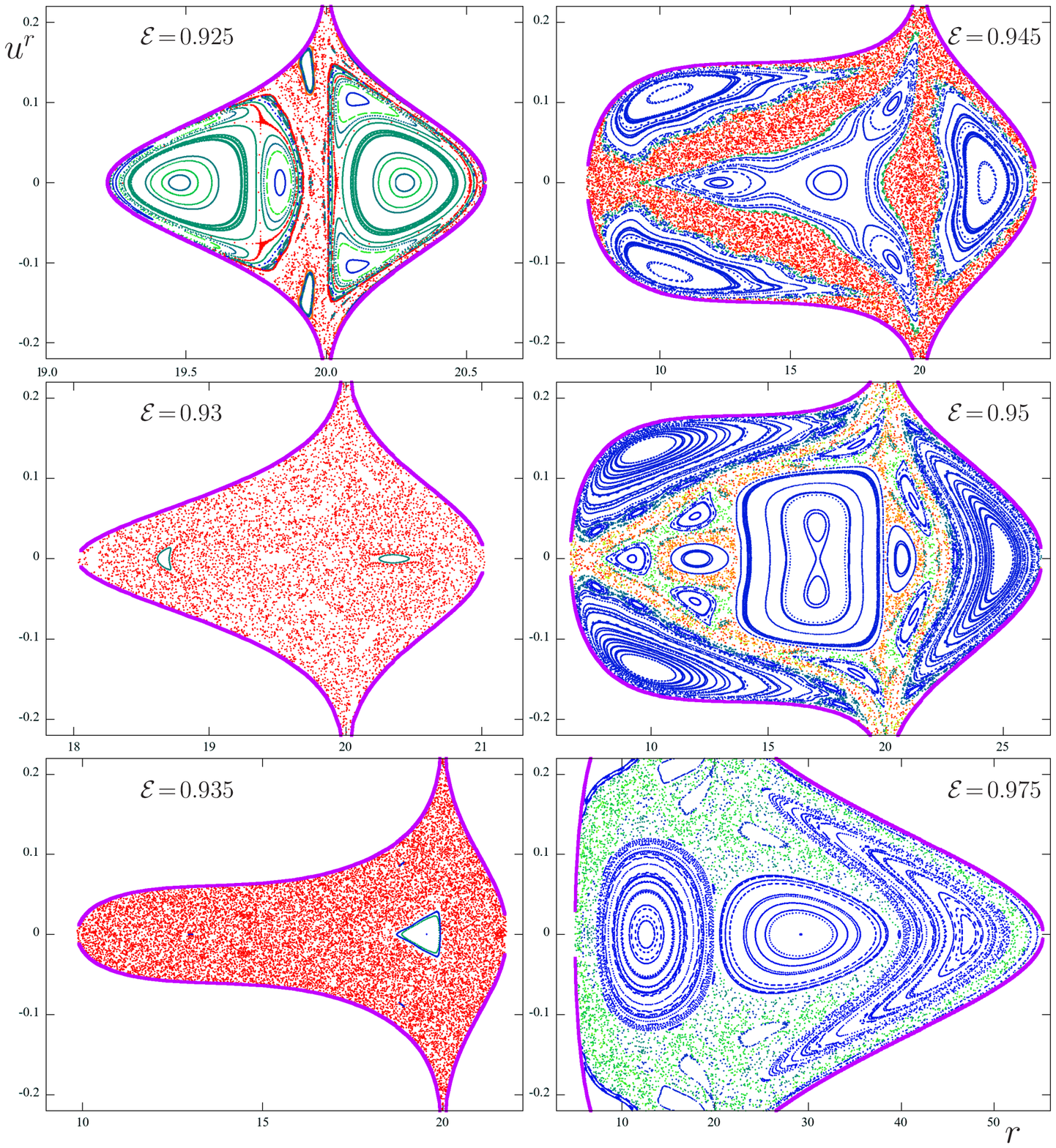}
\caption
{Counterpart of Fig. \ref{Schw-BW-energy} obtained for geodesics in the field of an extreme Reissner-Nordstr\"om black hole encircled by a Majumdar-Papapetrou ring. Parameters ${\cal M}$, $\ell$ and $r_{\rm ring}$ are chosen the same as in the Schwarzschild + Bach-Weyl ring case, while the sequence of energies ${\cal E}$ (indicated in the plots) is slightly different (namely, 0.925, 0.93, 0.935, 0.945, 0.95, 0.975, instead of 0.92, 0.925, 0.93, 0.94, 0.95, 0.97).}
\label{RN-MP-energy}
\end{figure*}

We will first examine, on Poincar\'e sections, whether the dynamics of bound time-like geodesics in the field of the RN black hole encircled by the MP ring differs significantly from the dynamics in the field of the Schwarzschild black hole encircled by the BW ring, studied in \citep{SemerakS-10}. The Poincar\'e diagrams represent sets of transitions of the particles across the equatorial plane in the $(r,u^r)$ axes. Each diagram corresponds to some particular parameters of the source (mass and radius of the ring/disc, the black-hole mass representing just a mass scale) and some particular values of the constant of geodesic motion; all geodesics followed in any diagram start in the equatorial plane, just from different $r$ and $u^r$ (the mesh of initial values being chosen so that the diagram be reasonably filled).

In order to make the Poincar\'e diagrams more explicative, we colour the orbits according to their value of MEGNO -- one of the simplest Lyapunov-type indicators, quantifying the rate of orbital divergence. It was explained and employed in \cite{SukovaS-13}, so let us just briefly repeat that it is defined by 
\begin{equation}  \label{MEGNO_def}
  Y(\tau) = 2\,[{\rm FLI}(\tau)-\overline{\rm FLI}(\tau)]\,\ln(10) \,,
\end{equation}
where the fast Lyapunov indicator (FLI) and its time average are calculated from the norm of a separation vector $\Delta x^\mu$ between two neighbouring orbits in the configuration space (i.e. from their momentary proper distance),
\begin{align}
  {\rm FLI}(\tau) &= \log_{10} \frac{\sqrt{|g_{\mu\nu}\Delta x^\mu\Delta x^\nu|(\tau)}}
                                    {\sqrt{|g_{\mu\nu}\Delta x^\mu\Delta x^\nu|(0)}} \;, \\
  \overline{\rm FLI}(\tau) &= \frac{1}{\tau}\int_0^\tau {\rm FLI}(s)\,{\rm d}s \,,
\end{align}
$\tau$ being proper time. Also often practical is the time-average value of MEGNO,
\begin{equation}  \label{prum_MEGNO}
  \overline{Y}(\tau)=\frac{1}{\tau}\int_0^\tau Y(s)\,{\rm d}s \,.
\end{equation}
The main advantage of MEGNO is that in the limit of very long proper time, it approaches the value of 2 for every quasi-periodic regular orbit (or a somewhat higher value for orbits which are regular but pass very close to unstable phase-space structures like resonances or hyperbolic points), whereas it grows linearly for chaotic orbits. It is thus very helpful in distinguishing between regular and chaotic trajectories.

For an easy comparison of the new (RN+MP) figures with the old (Schw+BW) ones, we repeat some of the plots provided, for the latter configuration, in \cite{SemerakS-10}, because there they were drawn just black and white (here we colour them in the same way as the new plots obtained for the other couple of sources).
The Schwarzschild + BW ring case is thus covered by Figs. \ref{Schw-BW-mass}, \ref{Schw-BW-radius} and \ref{Schw-BW-energy}, while the Reissner-Nordstr\"om + MP ring case by Figs. \ref{RN-MP-mass}, \ref{RN-MP-radius} and \ref{RN-MP-energy}, respectively.

\begin{figure*}
\includegraphics[width=\textwidth]{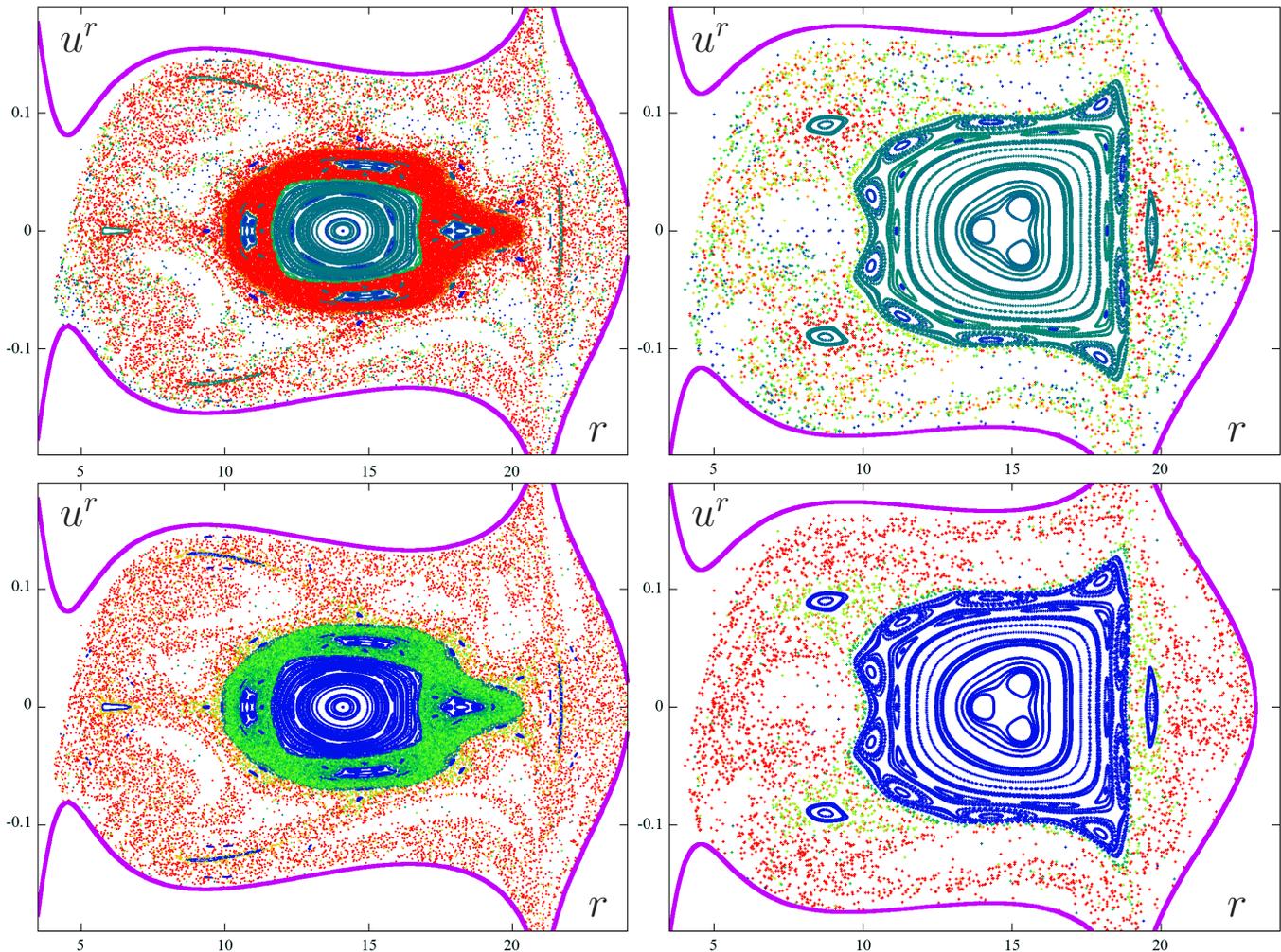}
\caption
{Two illustrations of a difference between MEGNO and what we call ``relative MEGNO'' (MEGNO achieved by a given orbit divided by proper time for which the orbit has been followed). Both columns of Poincar\'e diagrams show the dynamics of geodesics with constants of motion ${\cal E}\!=\!0.94$, $\ell\!=\!3.75M$ in the field of a Schwarzschild black hole encircled by the Bach-Weyl ring of mass ${\cal M}\!=\!0.5M$; in the left/right column, the ring has radius $r_{\rm ring}\!=\!19M\,/\,21M$, respectively. The top row is coloured by MEGNO (dark blue to red corresponds to its ranging from ``zero'' to more than 30), while the bottom row is coloured by relative MEGNO (ranging, through the same colour spectrum, from zero to more than 0.0008); no additional shift is applied. The colouring is strongly dependent on where one places its limits, but, according to our experience, the relative MEGNO more surely distinguishes between the regular and chaotic trajectories: in the top row, what is clearly a chaotic sea still contains many regular-looking (green or even blue) points, while, on the other hand, the regular island is not as strictly blue as in the bottom (despite trying to adjust the colouring suitably). Also note that the orbits which stick to the regular centre in the left case are assessed more adequately in the bottom plot, namely as being less chaotic than the chaotic sea.}
\label{MEGNOs-comparison}
\end{figure*}

The figures indicate that qualitatively the geodesic dynamics in both the fields does not differ much, the main difference actually being that in the case of the RN + MP ring the phase-space region where a particle with given energy and angular momentum can exist (it is indicated by the violet boundary in the figures) tends to be closed, whereas in the case of the Schw + BW ring it often opens towards the horizon, so in this latter case some particles fall to the black hole and only partially contribute to the diagram. For such trajectories, it is difficult to determine the degree of their chaoticity, because the indicators of orbital divergence (like MEGNO) only approach their relevant values at asymptotic times. For this reason, we have rather employed a ``relative MEGNO'' -- the MEGNO divided by proper time for which a given trajectory had been followed. Such a modified parameter has turned to better indicate the degree of chaoticity. Namely, for chaotic trajectories MEGNO typically grows linearly with time, so if a chaotic particle escapes from the system (e.g. falls to the black hole) too soon, its MEGNO reaches lower value than what would correspond to its actual nature. In our case this is particularly important, because when the accessible region opens towards the black hole, it is first left by particles from its outer parts, and exactly these are typically the most chaotic (``chaotic sea''). See Fig. \ref{MEGNOs-comparison} for two examples of how the relative MEGNO is a more precise indicator than the original MEGNO itself.

The colour scale employed, for the relative-MEGNO value, in all the figures ranges from 0 to 0.0005 (blue to red, in the spectral order), which means, roughly speaking, that blue and possibly green indicate regular orbits, while red marks the most chaotic ones.

\section{Numerical check of the Sota-Suzuki-Maeda curvature criterion}
\label{Sota-criterion,numerics}

\begin{figure*}
\includegraphics[width=\textwidth]{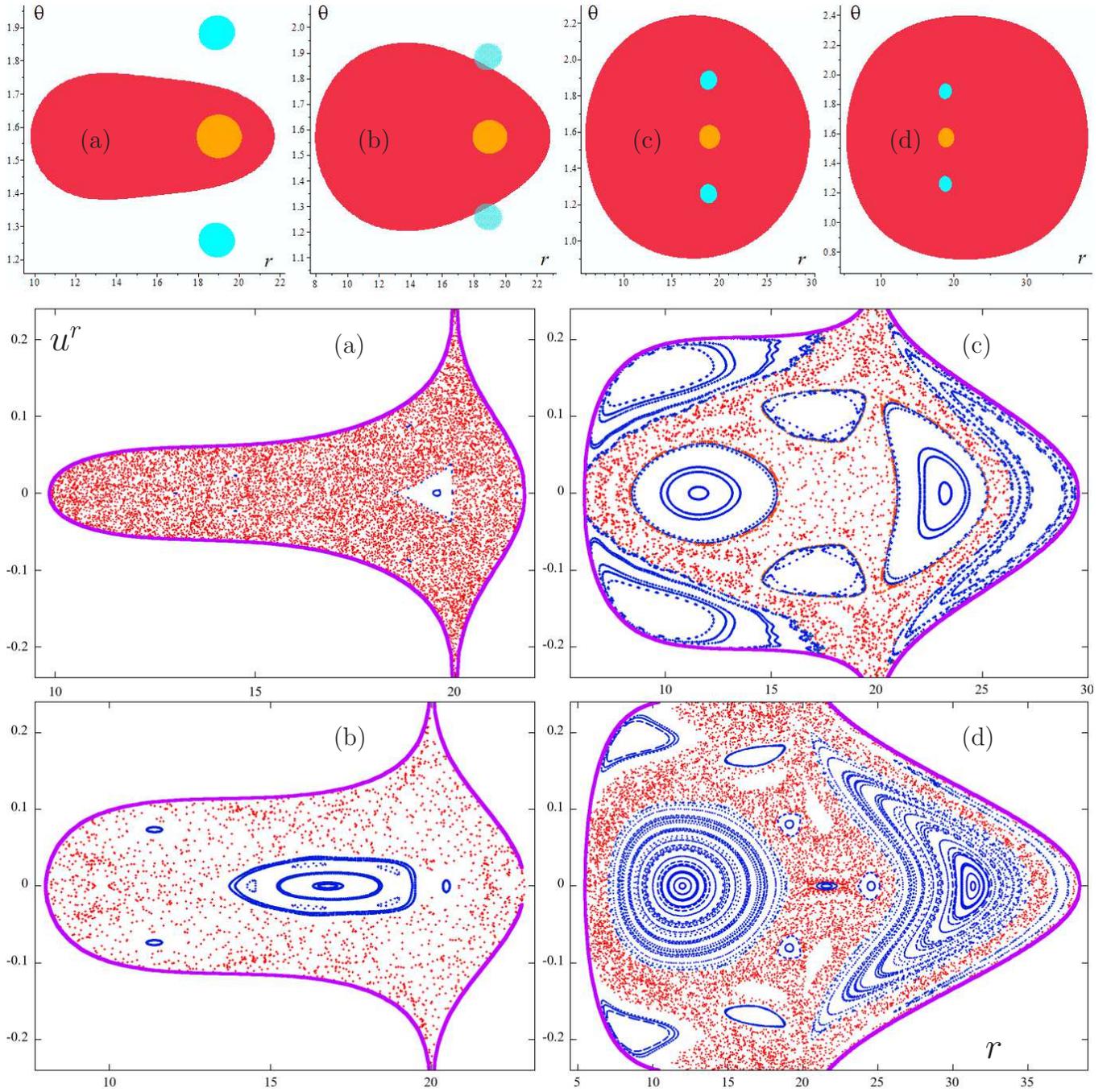}
\caption
{Comparison of the geometric criterion with the actual geodesic dynamics portrayed on Poincar\'e diagrams, performed, for four different particle energies, in the field of a Reissner-Nordstr\"om black hole encircled by the Majumdar-Papapetrou ring with mass ${\cal M}\!=\!0.5M$ and radius $r_{\rm ring}\!=\!20M$. All the geodesics have $\ell\!=\!3.75M$, with energies (a) ${\cal E}\!=\!0.935$, (b) ${\cal E}\!=\!0.940$, (c) ${\cal E}\!=\!0.955$, (d) ${\cal E}\!=\!0.965$.
{\it Top row:} Regions determined as unstable by the geometric criterion (plotted in the $r$, $\theta$ plane); red is the accessible region, light blue and orange are the unstable regions (with $+$$+$$-$ and $+$$-$$+$ signs of the curvature eigen-values, respectively).
{\it Bottom two rows:} Equatorial Poincar\'e diagrams of geodesics (plotted in the $r$, $u^r$ plane); transition points are coloured by MEGNO -- dark blue are regular orbits and red are chaotic orbits.}
\label{RN-MP-geom}
\end{figure*}

\begin{figure*}
\includegraphics[width=\textwidth]{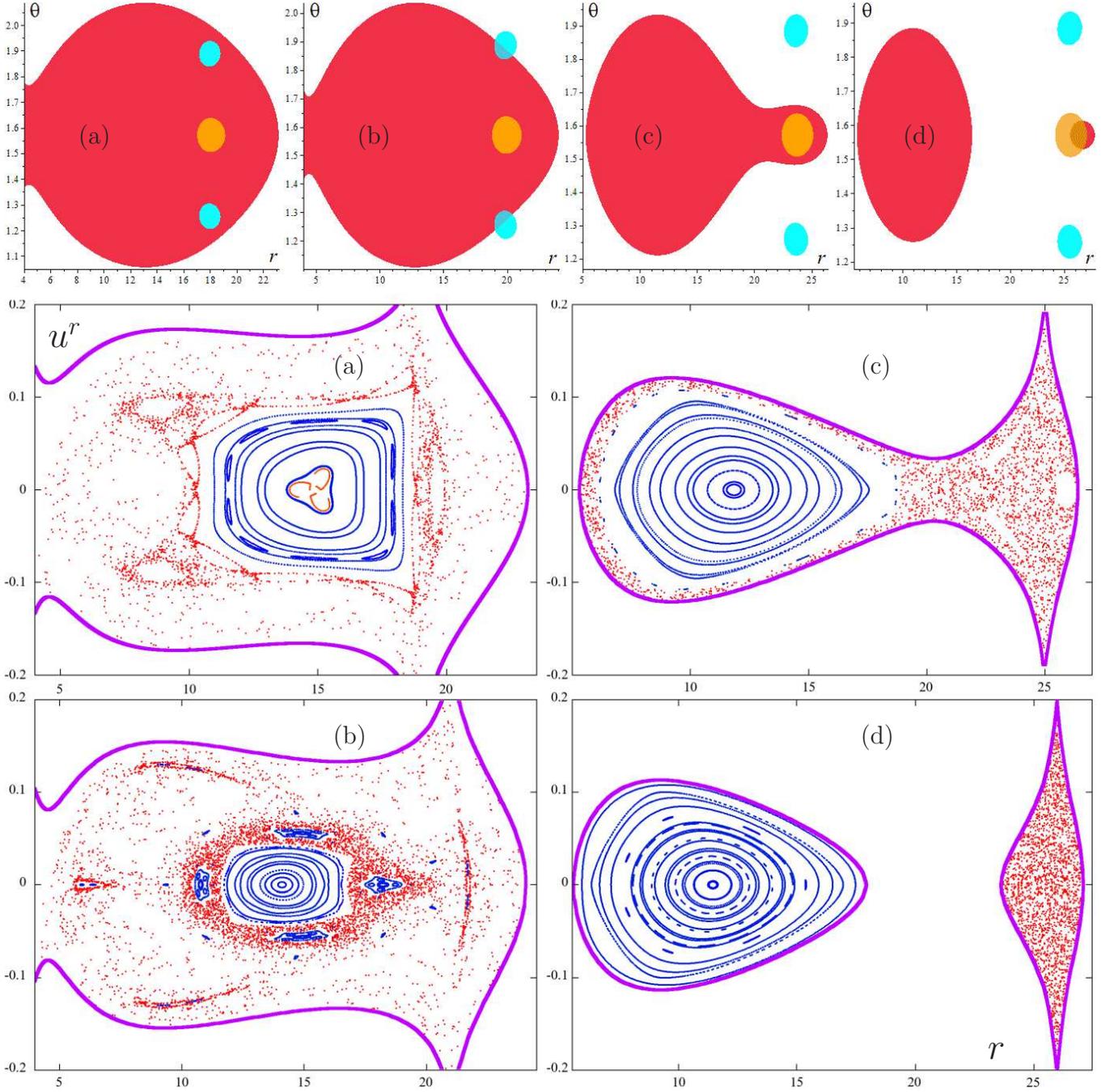}
\caption
{Comparison of the geometric criterion with the actual geodesic dynamics portrayed on Poincar\'e diagrams, performed in the field of a Schwarzschild black hole encircled by the Bach-Weyl ring with mass ${\cal M}\!=\!0.5M$ and with four different radii: (a) $r_{\rm ring}\!=\!19M$, (b) $r_{\rm ring}\!=\!21M$, (c) $r_{\rm ring}\!=\!25M$, (d) $r_{\rm ring}\!=\!27M$. All the geodesics have $\ell\!=\!3.75M$ and ${\cal E}\!=\!0.94$.
{\it Top row:} Regions determined as unstable by the geometric criterion (plotted in the $r$, $\theta$ plane); red is the accessible region, light blue and orange are the unstable regions (with $+$$+$$-$ and $+$$-$$+$ signs of the curvature eigen-values, respectively).
{\it Bottom two rows:} Equatorial Poincar\'e diagrams of geodesics (plotted in the $r$, $u^r$ plane); transition points are coloured by MEGNO -- dark blue are regular orbits and red are chaotic orbits.}
\label{Schw-BW-geom}
\end{figure*}

\begin{figure*}
\includegraphics[width=\textwidth]{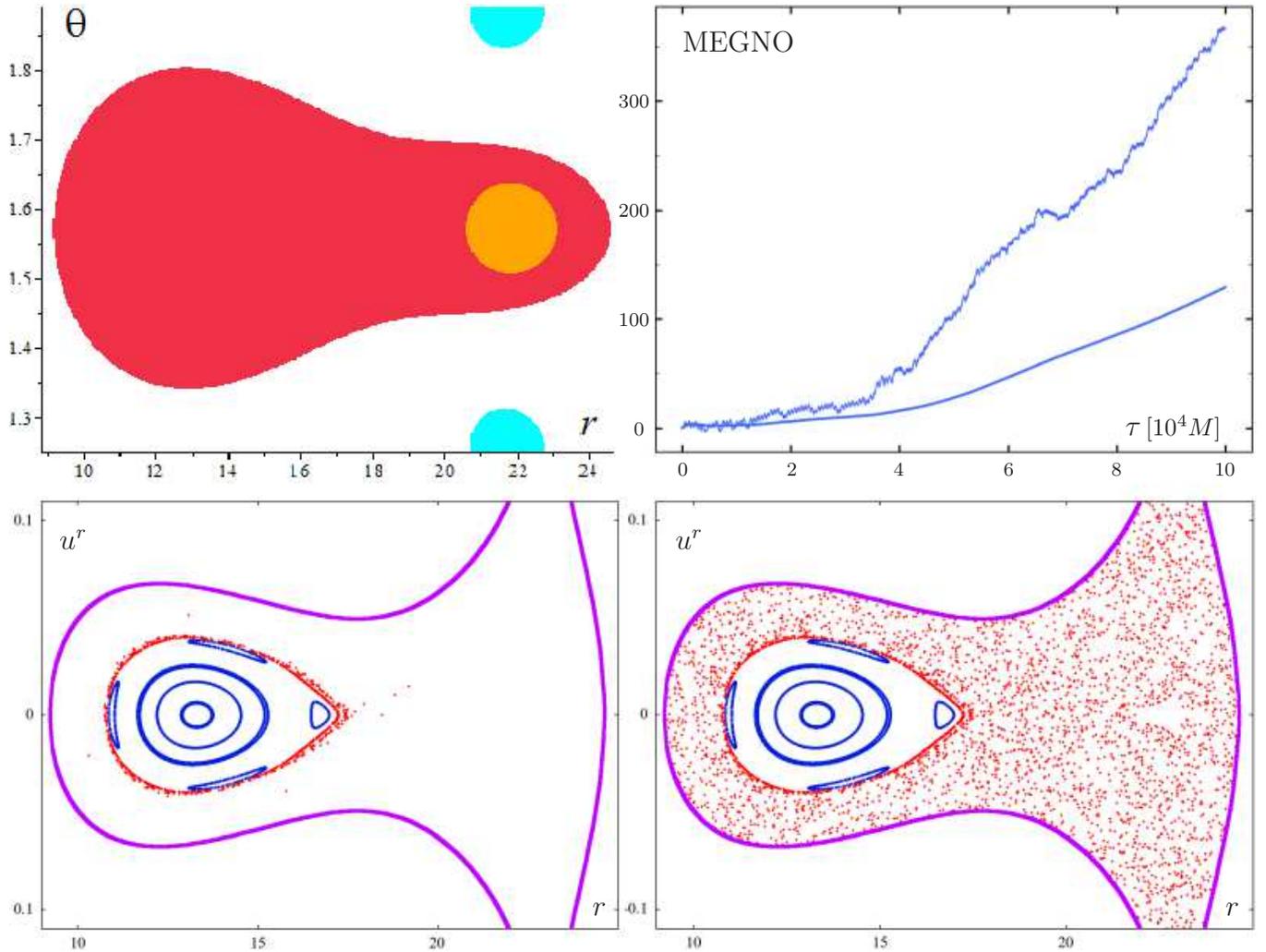}
\caption
{Example of a situation where the geometric criterion works very well: an orbit which sticks to a large regular island for quite some time, but then it begins to cross the ``orange'' unstable region (as identified by the criterion) and fills the whole accessible zone in a chaotic manner. It happens in the field of the extreme Reissner-Nordstr\"om black hole encircled, at $r_{\rm ring}\!=\!23M$, by the Majumdar-Papapetrou ring of mass ${\cal M}\!=\!0.5M$. The plotted geodesics have ${\cal E}\!=\!0.94$ and $\ell\!=\!3.75M$. Poincar\'e diagrams with a regular island (blue) and the chaotic orbit (red) are at the bottom -- the left one captures the first $3.5\cdot 10^4M$ of that orbit's proper time and the right one covers about $10\cdot 10^4M$ of proper time. The orbit turns from ``sticky'' to strongly chaotic after hitting the unstable region (the orange one at top left, lying within the red accessible region) at $\tau\!\simeq\!3.5\!\cdot\!10^4M$; this is clearly seen, at top right, on the increase of a slope with which MEGNO of the orbit grows (the bottom curve there shows the mean MEGNO).} 
\label{flewover}
\end{figure*}

\begin{figure*}
\includegraphics[width=\textwidth]{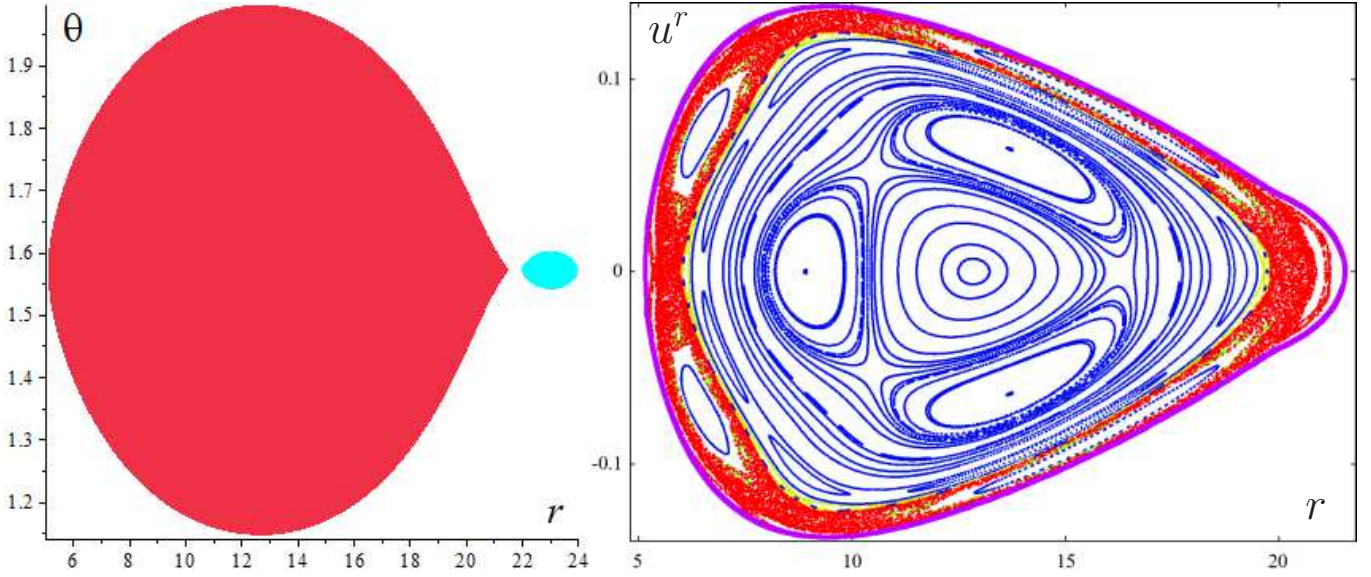}
\caption
{Example of a situation when the curvature criterion does not ``predict'' the dynamics properly: geodesics around a Schwarzschild black hole encircled, at $r_{\rm disc}\!=\!20M$, by the inverted first Morgan-Morgan disc of mass ${\cal M}\!=\!0.5M$. The geodesics have energy ${\cal E}\!=\!0.953$ and angular momentum $\ell\!=\!3.75M$. The left plot shows the only unstable region (light blue) {\em outside} of the accessible region (red), yet the Poincar\'e section on the right still reveals chaotic layer at the accessible-region boundary (region filled with red-coloured transitions).}
\label{Schw-iMM-geomfail}
\end{figure*}

\begin{figure*}
\includegraphics[width=\textwidth]{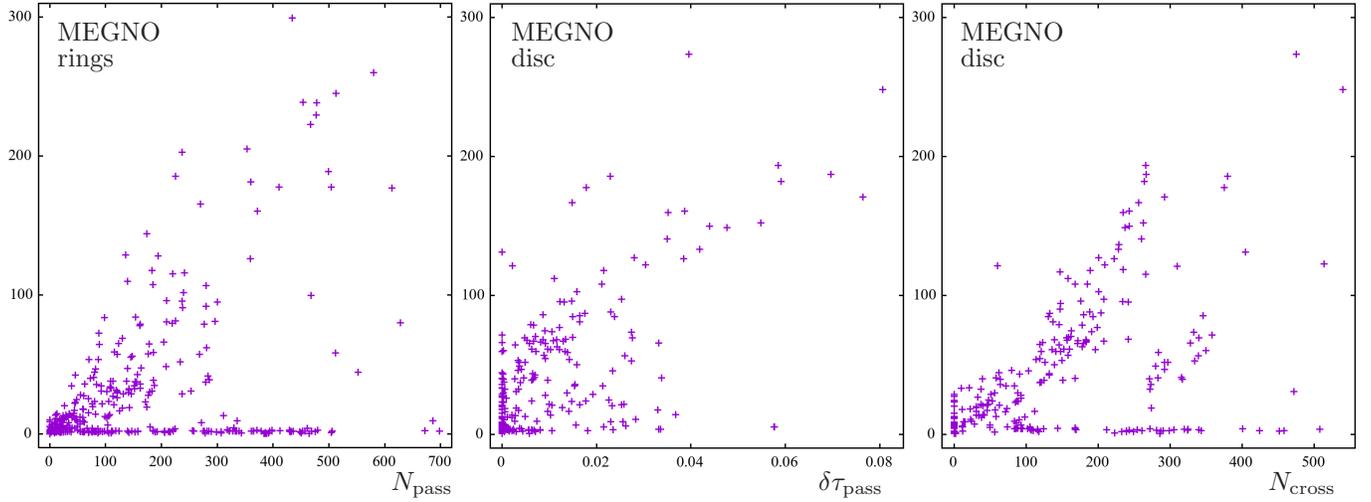}
\caption
{Terminal MEGNO of the orbits we followed, as plotted against simple numbers connected with possible destabilizing circumstances: in the left plot, it is the number of passages through unstable regions $N_{\rm pass}$ (included are results obtained for both ring superpositions we consider); the middle and right plots concern the inverted first Morgan-Morgan disc and MEGNO is plotted there against the relative proper time the orbits spent in the unstable regions ($\delta\tau_{\rm pass}$) and against how many times they crossed the disc ($N_{\rm cross}$). Apparently there is some correlation, but there are also many ``exceptions''.}
\label{correlations}
\end{figure*}

\begin{figure*}
\includegraphics[width=\textwidth]{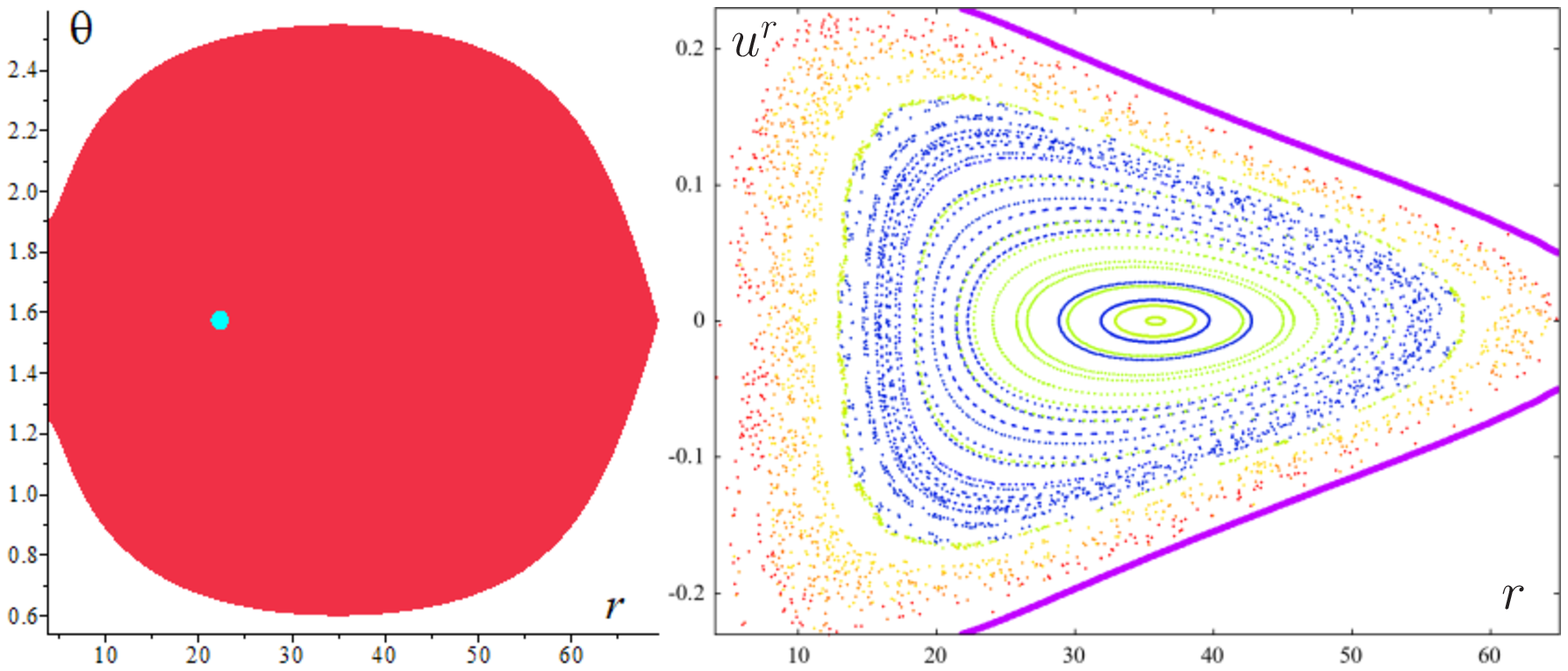}
\caption
{Example of a situation when neither the curvature criterion nor the incidence of disc crossings suggest proper picture of the dynamics: geodesics (with energy ${\cal E}\!=\!0.98$ and angular momentum $\ell\!=\!3.75M$) around a Schwarzschild black hole encircled, at $r_{\rm disc}\!=\!19.493M$, by the inverted first Morgan-Morgan disc of mass ${\cal M}\!=\!0.5M$. The left plot shows that the unstable region (light blue) does occur inside the accessible region (red), but it is so tiny that the computed geodesics practically do not cross it, yet the Poincar\'e section on the right still reveals wide chaotic layer at the outskirts of the accessible region (yellow-to-red coloured transitions). On the other hand, almost all the orbits crossed the disc repeatedly, yet there is still a notable regular island (blue-and-green coloured transitions).}
\label{both-fail}
\end{figure*}

We will naturally check the relevance of the Sota-Suzuki-Maeda criterion by computing phase-space portraits of the geodesic system (as recorded on Poincar\'e sections again) for different parameters and by comparing them with the location of the ``diverging'' regions identified by eigen-values of the tidal matrix determined by the Riemann tensor.
Note that in the following figures we draw the accessible region in red and the unstable regions given by the geometric criterion in light blue (with $+$$+$$-$ signs of the curvature eigen-values) and in orange (with $+$$-$$+$ signs).

Fig. \ref{RN-MP-geom} shows the comparison between the unstable-region maps given by the geometric criterion and Poincar\'e sections for geodesics in the field of the RN black hole encircled by the MP ring. The plots are given for several different values of the geodesic specific energy ${\cal E}$. For low energies the unstable regions are seen to occupy most of the accessible region, which should indicate rather strong inclination to chaos; with increasing energy, the unstable regions remain the same, while the accessible region grows, indicating attenuation of chaoticity. The Poincar\'e diagrams confirm such a tendency, though perhaps not as clearly as one could expect.

Fig. \ref{Schw-BW-geom} shows the same comparison for geodesics in the field of the Schwarzschild black hole encircled by the BW ring, this time for several different radii of the ring. All the unstable regions are seen to first (for $r_{\rm ring}\!=\!19M$) lie completely inside the accessible region, later -- with increasing ring radius -- the blue regions leave the accessible region; at some point, the accessible region spits into two, the lower one (closer to the black hole) completely devoid of unstable regions, whereas the upper one (located below the ring) almost ``covered by'' the orange unstable region. This nicely agrees with what one sees in Poincar\'e's diagrams, in particular, the last one ($r_{\rm ring}\!=\!27M$) shows a completely regular lower part of the accessible region and a completely chaotic upper part. A similar observations can be inferred from a series showing dependence on the ring mass (not shown here).

In previous papers of this series, we also considered, around a Schwarzschild black hole, a thin equatorial disc, specifically the one obtained by inversion of some member of the Morgan-Morgan counter-rotating family. We have now also tested the geometric criterion in such a gravitational background and have {\em not} found the agreement with the actual dynamics as good as for the rings. This may seem surprising since the discs, being spatially two-dimensional, are more physical (as relativistic sources) than the one-dimensional rings, so intuition can be expected to work better for them. However, what is probably disturbing are the transits of particles across the disc, because the field jumps across the disc in the perpendicular direction, like across any source layer.\footnote
{If the particles hit the ring or come to its closest vicinity, their motion is just terminated, because the ring represents a curvature singularity and, therefore, has to be ``excised'' by the code. On the other hand, the disc is regular and there is no reason to stop the particles at it (and we neglect any possible mechanical interaction) -- in fact they {\em typically} cross the disc repeatedly at some stage of their motion.}
As an example of situation where the criterion does not work well, see Fig. \ref{Schw-iMM-geomfail}. There, no unstable regions occur within the accessible lobe, yet the Poincar\'e diagram still contains chaotic layers.

\subsection{Quantification of the geometric-criterion effectiveness}

The locations of unstable regions and the corresponding Poincar\'e sections provide only a summary picture of what should in fact arise from {\em local} behaviour (restricted to certain spatial regions and thus happening only during certain intervals of time). In order to at least roughly quantify the comparison, it is desirable to track down whether, when, for how long or at least how often a {\em specific} particle has crossed the unstable region and, on the other hand, how the chaos-indicating parameters has evolved along its trajectory.
For this purpose, we will compute the time evolution of MEGNOs for some particular orbits and, on the other hand, record how much of their proper time these orbits spent in crossing the ``diverging'' regions determined by curvature (or/also how many times they have crossed such regions). The degree of correlation between these quantities could indicate whether the criterion is effective or not. One might of course consider a more sophisticated quantity like an average of the $\lambda_1$ and $\lambda_2$ values along the orbit, but we will adhere to the former simple possibilities.

It is not any problem to find either regular or chaotic orbits for which the geometric criterion works very well. However, most interesting are such orbits which start in a ``regular'' region but, after some time, get to a region where the geometric criterion indicates instability (and then, possibly, leave to the regular part of phase space again). An example of such an orbit is given in Fig. \ref{flewover}: the figure contains four plots, the first showing location of unstable regions within the accessible part of phase space, the second showing evolution of MEGNO with proper time, and the third and fourth showing Poincar\'e sections recorded {\em before} hitting the unstable region for the first time (left) and recorded for the whole trajectory (right). The geometric criterion works perfectly here, namely the orbit is almost regular before and quite chaotic after it hits the unstable region, with the MEGNO slope really increased at time when the unstable region is first entered.

However, there exist orbits whose character the criterion does not predict correctly. For example, we noticed that this can be the case for orbits which are regular, but lie close to a resonant or hyperbolic torus (this is typically indicated by MEGNO going asymptotically to a constant value larger than 2, see \citealt{Maffione-etal-11}). These orbits may spend quite some time in an unstable region, so the geometric criterion wrongly predicts their chaotic nature. To give a specific example, we found a regular orbit with asymptotic MEGNO of 3.054, for which the relative proper time spent in an unstable region was 0.0323. (This value may seem small, but we should add that even the most chaotic orbits we studied spent just several percent of their proper time in unstable regions. See the middle plot of Fig. \ref{correlations} where MEGNO is plotted against total relative time the orbits spent in the unstable regions, so one gets an idea from there about typical values.) The above experience may indicate that unstable structures of the phase space (hyperbolic orbits) are not correlated with the presence of unstable regions given by curvature.

Another piece of information on effectiveness of the geometric criterion can be provided by plotting, for a large number of orbits, the terminal value of MEGNO achieved in a simulation against the number of passages through the unstable region(s) determined by the criterion -- see Fig. \ref{correlations} where such a plot is shown (it is the left one there) containing orbits around both centres considered above (RN black hole encircled by MP ring and Schwarzschild hole encircled by BW ring); a similar result can be obtained for correlation between the MEGNO and the relative proper time the orbits spend in the unstable regions. In general, MEGNO really grows with the number of passages, but there exist orbits with many passages yet with MEGNO remaining small (the points along the horizontal axis). These, in particular, include the orbits mentioned in the previous paragraph. The figure -- and our experience in general -- can be summarized so that the geometric criterion seems to be necessary, but not sufficient for the geodesic dynamics to be chaotic (there are no orbits clustered along the vertical axis).

Turning now to the case of a Schwarzschild black hole encircled by a thin disc (the inverted first Morgan-Morgan disc in particular), we saw that for this gravitational background the geometric criterion even fails to be a necessary condition for chaos (see example in Fig. \ref{Schw-iMM-geomfail}). In order to test our conjecture that it may be due to the crossings of the disc by the geodesics, let us check a different correlation, between the achieved value of MEGNO and the number of times it passed across the disc. Fig. \ref{correlations} (right two plots) confirms that such a correlation really turns out to work better (like, for example, in the situation represented in Fig. \ref{Schw-iMM-geomfail} where the curvature criterion failed). Yet neither this correlation is always valid -- for example, in Fig. \ref{both-fail} almost all geodesics cross the disc many times, but the Poincar\'e section still contains a large regular island. In the left plot of that figure, ``prediction'' of the curvature criterion is also showed: a tiny blue unstable region lies inside the accessible region, but is almost never hit by any of the particles (yet there is a chaotic orbit at the outer parts of the accessible region, with the reached value of MEGNO equal to 31.11).

As a certain summary of results obtained for the Schwarzschild black hole encircled by the inverted first Morgan-Morgan disc, we again plot, for several tens of orbits, their terminal values of MEGNO against the relative proper time spent in regions predicted as unstable (middle plot of Fig. \ref{correlations}), but also against the number of crossings through the disc (right plot of Fig. \ref{correlations}). The correlation is clearly stronger for the second plot, but even there one sees cases with many crossings yet low MEGNO. The thin-disc case would thus require a more detailed study, but it is possible that neither that would lead to any clear conclusion, as also experienced in the literature. Let us add, on the other hand, that we have {\em not} encountered the case when the unstable region(s) would fill almost the whole accessible lobe yet the dynamics would still be completely regular (this would be the strongest counter-example to the geometric criterion).

\section{Concluding remarks}

A full geodesic integrability is one of remarkable features of space-times admitting (or actually ``generated by'' in a mathematical sense) a non-degenerate closed conformal Killing-Yano 2-form, as recently summarized thoroughly by \cite{FrolovKK-17}. Such space-times are called to possess ``hidden symmetries''; they are necessarily of curvature type D and represent, essentially, those of isolated stationary black holes (loosely speaking, those of the Kerr type).\footnote
{It should be specified that the black holes have to be non-accelerating. Actually, there exist black-hole type-D space-times which do {\em not} possess hidden symmetries, and thus their geodesics are in general not completely integrable. They are described by ``C-metric--type'' solutions (subclass of the Pleba\'nski-Demia\'nski metrics) and in the accelerating case their geodesics can really show chaotic behaviour (see \citealt{ChenWJ-16}).}
The integrability is ensured by the existence of the ``fourth'' integral of (electro-)geodesic motion (besides the momentum norm and the integrals following from stationarity and axisymmetry); in a Newtonian case, its counter-part represents the ``third'' integral and its existence was e.g. investigated by \cite{Henon-Heiles-64}.

In the present paper, we continued to study how a deviation from the above Kerr-like ideal destabilizes the geodesic dynamics. Restricting to the simple case of spherically symmetric black holes, we considered their perturbations due to a thin ring or disc, specifically, the Bach-Weyl ring around a Schwarzschild black hole, the Majumdar-Papapetrou (extremely charged) ring around an extreme Reissner-Nordstr\"om black hole, and the inverted first disc of the counter-rotating Morgan-Morgan family around the Schwarzschild black hole. One new point has been the inclusion of the electrically charged case, motivated by the much more reasonable behaviour of geometry in the vicinity of the MP ring (than around the BW ring). In spite of quite a different nature of space-time in that case (it is no longer vacuum), the geodesic dynamics appears to undergo, with growing strength of the perturbation, similar stages like in the vacuum, Schwarzschild + BW case. More profound differences could rather be expected if the additional source broke the reflection symmetry.

Let us mention some recent publications on chaotic motion around perturbed black holes. \cite{KopacekK-14} analysed perturbation exerted on charged particles by an external large-scale magnetic field inclined with respect to the black-hole rotation axis. \cite{LiuWH-17} studied test motion around Schwarzschild perturbed by a shell of dipoles, quadrupoles or octupoles, and \cite{LiWu-18} considered the Schwarzschild black hole immersed in a magnetic field. \cite{NagSAD-17} observed, within a pseudo-Newtonian treatment, that the space-time dragging due to the centre's (black-hole) rotation has an attenuating effect on geodesic chaos. \cite{LG-Kopacek-18} used recurrence analysis for a particle in-spiralling in a deformed Kerr field while emitting gravitational waves, and showed that the character of motion can be recognized even if noise is present in the observed signal. Let us also mention \cite{Bannikova-18} who studied the motion in the field of a torus (as approximated by a thin ring), inspired by the case of ring galaxies (Hoag's objects).

Another our point has been to check the validity of one of the curvature-based estimates of chaos, due to \cite{SotaSM-96}. We confirmed that it is neither necessary nor sufficient, although it mostly works as a useful indicator.
It should be admitted here that our fields contain singularities (thin sources themselves, namely rings or discs) which is {\em not} a situation where one would guess the geometric criteria for geodesic chaos could work reliably, so next we might either focus on motion kept away from these irregularities, or consider more regular (extended) sources, e.g. a thick toroid instead of the thin ring. The relation between thick toroids and their infinitesimally thin ring limits is definitely worth a further study, and similarly is the influence of field irregularities caused by thin matter configurations, mainly in order to judge how major an error one introduces when employing such sources as approximation of real astrophysical bodies. This does not only concern infinitesimally thin rings (we have a clear experience that the ``contact'' of particles with such a ring is strongly destabilizing), but also razor-thin layers (discs) which generate jump in the normal field. As shown by \cite{VieiraRC-16} within a Newtonian treatment, in the latter case the destabilization need not be that strong, namely for orbits crossing such a thin disc there still typically exists an approximate ``third'' integral of motion (besides energy and angular momentum).

Other plans for future include using still another methods like that of Melnikov's integral or the basin-boundary analysis, and, needless to say, more interesting (and astrophysically more adequate) space-times involving rotation.

\acknowledgments

We thank for support from the grants GACR-17/06962Y (L.P., P.S.) and GACR-17/13525S (O.S.) of the Czech Science Foundation.
We also acknowledge that our numerics has still been based on a code written two decades ago by one of O.S.'s students M. \v{Z}\'a\v{c}ek.

\end{document}